\documentclass[journal]{IEEEtran}
\usepackage{hyperref}
\usepackage{color}
\usepackage{amsmath}
\usepackage{bm}
\usepackage{mathrsfs}
\usepackage{multirow}
\usepackage{subfig}
\usepackage{cite}
\usepackage{graphicx}
\usepackage{amssymb}
\usepackage{calc}
\usepackage{stfloats}
\usepackage{comment}
\usepackage{textcomp,gensymb,cite}
\usepackage{booktabs}
\usepackage{array}
\usepackage{tabularx}
\usepackage{epstopdf}
\usepackage{multirow}
\usepackage{threeparttable}
\graphicspath{{Figure/}}

\newtheorem{remark}{Remark}

\usepackage[linesnumbered,ruled,vlined]{algorithm2e}

\SetCommentSty{mycommfont}

\SetKwInput{KwInput}{Input}                % Set the Input
\SetKwInput{KwOutput}{Output}              % set the Output

\newcounter{MYtempeqncnt1}
\newcounter{MYtempeqncnt2}
\newcounter{MYtempeqncnt3}
\newcounter{MYtempeqncnt4}
\newcounter{MYtempeqncnt5}
\newcounter{MYtempeqncnt6}
\newcounter{MYtempeqncnt7}

\makeatletter
\def\endthebibliography{%
  \def\@noitemerr{\@latex@warning{Empty `thebibliography' environment}}%
  \endlist
}

\begin{document}

\title{Joint State Estimation and Noise Identification Based on Variational Optimization}

%\title{Adaptive Kalman Filtering Based on Conjugate-computation Variational Inference}

\author{Hua Lan\thanks{Hua Lan, Shijie Zhao, Jinjie Hu and Zengfu Wang are with the School of Automation, Northwestern Polytechnical University, Xi'an, Shaanxi, 710072, P.~R.~China.
Jing Fu is with School of Engineering, RMIT University, Melbourne, VIC, 3000, Australia.
This work was supported in part by the National Natural Science Foundation of China (Grant No.~62371398, U21B2008) Corresponding author:~Zengfu Wang~(wangzengfu@nwpu.edu.cn).}, Shijie Zhao, Jinjie Hu, Zengfu Wang, Jing Fu}
\maketitle
\begin{abstract}
%Most state estimation methods require knowledge of the dynamical models and the noise statistics. However, in many real applications, e.g., maneuvering target tracking, the noise statistics are often unknown or only partially known. Thus, noise parameter identification is an essential part of adaptive Kalman filtering. 
In this article, the state estimation problems with unknown process noise and measurement noise covariances for both linear and nonlinear systems are considered. By formulating the joint estimation of system state and noise parameters into an optimization problem, a novel adaptive Kalman filter method based on conjugate-computation variational inference, referred to as CVIAKF, is proposed to approximate the joint posterior probability density function of the latent variables~(i.e., the noise covariance matrices and system state). 
Unlike the existing adaptive Kalman filter methods utilizing variational inference in natural-parameter space, CVIAKF performs optimization in expectation-parameter space, resulting in a faster and simpler solution, particularly for nonlinear state estimation. Meanwhile, CVIAKF divides optimization objectives into conjugate and non-conjugate parts of nonlinear dynamical models, whereas conjugate computations and stochastic mirror-descent are applied, respectively. Remarkably, the reparameterization trick is used to reduce the variance of stochastic gradients of the non-conjugate parts. 
The effectiveness of CVIAKF is validated through synthetic and real-world datasets of maneuvering target tracking.
\end{abstract}

\begin{IEEEkeywords}
Nonlinear state estimation, noise covariance identification, adaptive Kalman filtering, computation-conjugate variational inference, maneuvering target tracking
\end{IEEEkeywords}

\section{Introduction}
Most state estimation methods are based on the assumptions that the dynamical models and the noise covariance matrices of the studied system are known \emph{a priori}. 
However, in practice, obtaining accurate statistics of the noise can be difficult or even impossible. For example, in noncooperative target tracking~\cite{gholson1977maneuvering}, the noise covariance matrices are often time-varying and unknown, arising from unexpected maneuvers of noncooperative targets and measurement outliers. In such cases, standard
Kalman filtering methods that rely on knowledge of noise statistic parameters may lose their accuracy or even lead to filter divergence. To solve this problem, one must resort to an adaptive Kalman filter~\cite{li2021state}, which performs state estimation and noise identification simultaneously.

There are two primary categories of adaptive Kalman filter methods, namely model adaptive methods and parameter adaptive methods. The model adaptive methods consist of the so-called multiple model estimators~\cite{li1996multiple}, which perform the joint state estimation and model selection by employing a series of candidate models characterizing different system modes. Optimal state estimation is intractable for multiple model systems due to the exponential growth of hypotheses. Therefore, some suboptimal approaches have been investigated to reduce the number of underlying hypotheses, including the well-known interactive multiple model estimator~(IMM)~\cite{blom1988interacting, mazor1998interacting}. The model adaptive methods show promising performance in maneuvering target tracking, signal processing, fault detection, etc. However, the requirement of available complete model sets restricts their capability of tackling model uncertainties. The parameter adaptive methods consist of the so-called equivalent-noise methods~\cite{li2002survey}, whereas the transition function of the dynamic system and the measurement function of the sensor are known, and the equivalent noise quantifies the modeling error, e.g., maneuvers. 
The noise statistics are nonstationary and unknown, which need to be identified along with the state estimation. The main idea of parameter adaptive methods is to infer the intractable joint posterior probability density function~(PDF) of system state and noise parameter. 
%The optimal Bayesian inference may not have closed-form analytical solutions. 
Closed-form analytical solutions may not exist for optimal Bayesian inference.
In such situations, one needs to resort to approximation schemes. There are mainly two categories of methods, including sampling-based stochastic approximation~\cite{arulampalam2002tutorial} and optimizing-based deterministic approximation, which trade off speed, accuracy, simplicity, and generality.
The sampling-based sequential Monte Carlo~(SMC)~\cite{arulampalam2002tutorial} methods approximate the intractable joint PDFs via particle propagation~\cite{ozkan2013marginalized}.
However, their applicability is restricted to small-scale state estimation problems due to the significant computational burden they impose.
%but are limited to small-scale state estimation problems due to the demanding computational burden. 
The optimizing-based variational Bayes~(VB)~\cite{Blei2016variational} reduces posterior inference into optimization, yielding an approximate analytical solution. Due to the efficient computation for high-dimensional estimation problems, VB-based methods have received ubiquitous interest in adaptive filtering applications.

Considering that the covariance matrix of the measurement noise is unknown, 
S\"{a}rkk\"{a} and Nummenmaa~\cite{sarkka2009recursive} presented the first VB-based adaptive Kalman filtering, 
whereas the Gaussian Inverse-Gamma variational distributions are used to approximate the joint PDFs of system state and noise parameters. This work was further extended to adaptive Kalman filtering~\cite{Huang2017TAC} and smoother~\cite{ardeshiri2015approximate} with both unknown process and measurement noise covariances. 
Ma~\emph{et al.}~\cite{ma2018multiple} approximated the joint PDF of the state and model identity via VB to solve the state estimation of multiple state-space models. 
Leveraging Student-t noise modeling, Zhu~\emph{et al.}~\cite{zhu2022sliding} proposed a variational outlier-robust Kalman filter by utilizing sliding window measurements to distinguish model uncertainties. 
Yu and Meng~\cite{yu2023robust} considered the joint estimation of state and multiplicative noise covariance and proposed VB-based robust Kalman filters. Xu~\emph{et al.}~\cite{xu2020adaptive} proposed the VB-based adaptive fixed-lag smoothing with unknown measurement noise covariance. Xia~\emph{et al.}~\cite{xia2021fine} proposed an adaptive variational Kalman filter with unknown measurement noise to solve the calibration problem. 
Zhu~\emph{et al.}~\cite{zhu2021novel, zhu2021adaptive} proposed variational Kalman filters designed for state estimation in the presence of unknown, time-varying, and non-stationary heavy-tailed process and measurement noises.

Most existing VB-based adaptive Kalman filter methods are limited to linear state space models, whereas the posterior PDFs of state and noise covariance are updated via the coordinate ascent under the requirement of conjugate prior PDF. 
For nonlinear state space models, the coordinate ascent loses its effectiveness in maximizing the evidence lower bound (ELBO). 
The common solution is combining the adaptive Kalman filter with nonlinear approximations, 
such as adaptive Metropolis sampling~\cite{2015MbalawateAdaptive} and 
cubature integration rule~\cite{2017DongVariational}. 
%For instance, S\"{a}rkk\"{a} and Hartikainen~\cite{2015MbalawateAdaptive} extended their work~\cite{sarkka2009recursive} with an . 
%Dong~\emph{et al.}~\cite{2017DongVariational} proposed an adaptive cubature information filter with the . 
Stochastic-gradient methods~\cite{hoffman2013stochastic} are an effective approach to tackle this issue. Recently, Lan~\emph{et al.}~\cite{10247583} proposed the nonlinear adaptive Kalman with unknown process noise covariance based on stochastic search variational inference, achieving high accuracy but suffering from slow iteration convergence. 
%An alternative approach is to use natural gradient descent~\cite{martens2020new}, which exploits the Riemannian geometry of variational distribution to facilitate faster convergence~\cite{hensman2012fast}, compared with the stochastic gradient descent. 
Another approach is to employ natural gradient descent~\cite{martens2020new}, leveraging the Riemannian geometry of the variational distribution for accelerated convergence compared to stochastic gradient descent~\cite{hensman2012fast}.
%Unfortunately, the simplicity of natural-gradient updates is limited to a restricted class of conditionally conjugate models and is not ineffective for non-conjugate models~\cite{lin2019fast}. 
Unfortunately, the simplicity of natural-gradient updates is limited to a specific class of conditionally conjugate models and proves ineffective for non-conjugate models, as indicated by Lin \emph{et al.}~\cite{lin2019fast}.
Khan and Lin~\cite{khan2017conjugate} proposed a conjugate-computation variational inference~(CVI) method, which generalizes the use of natural gradients to complex non-conjugate models. Different from the existing VB methods performing the natural-gradient descent in hyperparameter or natural-parameter space, the CVI method optimizes in expectation parameter spaces, enabling a simple and efficient natural-gradient update.

This article considers state estimation problems with unknown process and measurement noise covariances for linear and nonlinear systems. By formulating the joint estimation of system state and noise covariances into a variational optimization problem, a novel adaptive Kalman filter method derived from conjugate-computation variational inference, referred to as CVIAKF, is proposed to approximate the joint PDFs of latent variables. The CVIAKF uses a stochastic mirror-descent method in the expectation-parameter space which differs from the existing VB-based adaptive Kalman filters that perform in the natural-parameter space. For linear systems, the CVIAKF updates the variational hyperparameters in closed-loop solutions. 
For nonlinear systems, CVIAKF decomposes the ELBO optimization problem into conjugate terms and non-conjugate terms. Conjugate computations are employed for the former, while stochastic mirror gradients are applied for the latter.
Finally, the effectiveness of CVIAKF is verified on both synthetic and real-world datasets for maneuvering target tracking applications.

The remainder of the article is organized as follows. Section II elaborates on the problem formulation of the adaptive Kalman filter with unknown process noise covariance and measurement noise covariance. Section III presents the proposed CVIAKF method for both linear and nonlinear systems. 
Section IV and Section V evaluate the performance of CVIAKF by comparing it with the state-of-the-art adaptive Kalman filters using both synthetic and real-world datasets, respectively. 
Finally, the conclusion is given in Section VI.

\section{Problem Formation}
The discrete-time state space model is described as
\begin{equation}
\label{stochastic-systems}
\begin{split}
    \bm{x}_k &= \bm{f}_k\left( \bm{x}_{k-1} \right) +\bm{v}_{k},\\
    \bm{y}_k &= \bm{h}_k\left( \bm{x}_k \right) +\bm{w}_k,
\end{split}
\end{equation}
where $k$ is the time index; $\bm{x}_k$ and $\bm{y}_k$ are the state vector and measurement vector; $\bm{f}_k(\cdot)$ and $\bm{h}_k(\cdot)$ are the state transition function and measurement function; 
$\bm{v}_k \sim \bm{\mathrm{N}}(\bm{0}, \bm{Q}_k)$ is process noise,
$\bm{w}_k \sim \bm{\mathrm{N}}(\bm{0}, \bm{R}_k)$ is measurement noise,
the initial state $\bm{x}_0 \sim \bm{\mathrm{N}}(\hat{\bm{x}}_{0|0},\bm{P}_{0|0})$,
where $\bm{\mathrm{N}}(\cdot, \cdot)$ denotes a Gaussian distribution.
Moreover, $\bm{x}_0$, $\bm{v}_{k}$ and $\bm{w}_k$ are mutually independent. 

Under the assumptions of linear transition and observation functions $f_k$ and $h_k$, Gaussian process noise $\bm{v}_k$ and Gaussian observation noise $\bm{w}_k$, and the full knowledge of $\bm{Q}_k$ and $\bm{R}_k$, the Kalman filter achieves optimality with respect to minimizing the minimum mean square error~(MMSE) for state-space model~\eqref{stochastic-systems}. 
However, in many real applications, $\bm{Q}_k$ and $\bm{R}_k$ are often unknown or only partially known. 
For instance, in radar target tracking, $\bm{Q}_k$ is unknown since the maneuver of the noncooperative target is unpredictable. The covariance $\bm{R}_k$, correlated to the radar waveform and signal-to-noise ratio~\cite{skolnik1960theoretical}, is time-varying and uncertain. 
%Using wrong or inaccurate $\bm{Q}_k$ or $\bm{R}_k$ may result in substantial estimation errors or even filtering divergence.
Employing incorrect or imprecise $\bm{Q}_k$ or $\bm{R}_k$ values can lead to significant estimation errors or even filter divergence.

%The task of adaptive Kalman filtering is to perform the joint state estimation and noise covariance identification. 
In the Bayesian estimation framework, adaptive Kalman filtering requires to solve the joint posterior PDFs $p(\bm{\Theta}_k|\bm{y}_{1:k})$ of latent variables $\bm{\Theta}_k \triangleq \{\bm{x}_k, \bm{Q}_k, \bm{R}_k\}$, including system state $\bm{x}_k$ and noise covariance matrices $\bm{Q}_k$, $\bm{R}_k$. The optimal Bayesian filtering is required to recursively solve the predict-update cycle:
\begin{itemize}
  \item \emph{Initialization}: Initialize prior PDF $p(\bm{\Theta}_0)$.
  \item \emph{Prediction}: Calculate the predicted PDF of $\bm{\Theta}_k$ via the Chapman-Kolmogorov equation:
        \begin{equation}\label{eq::prediction}
            \begin{split}
                p(\bm{\Theta}_k|&\bm{y}_{1:k-1})  = \int p(\bm{x}_k|\bm{x}_{k-1}, \bm{Q}_k)p(\bm{Q}_k|\bm{Q}_{k-1}) \\ 
                &\times p(\bm{R}_k|\bm{R}_{k-1})p(\bm{\Theta}_{k-1}|\bm{y}_{1:k-1})\text{d}\bm{\Theta}_{k-1}.
            \end{split}
        \end{equation}
  \item \emph{Update}: Update $p(\bm{\Theta}_k|\bm{y}_{1:k-1})$  with measurement $\bm{y}_k$ according to the Bayes' rule:
            \begin{equation}\label{eq::update}
                p(\bm{\Theta}_k|\bm{y}_k) \propto p(\bm{y}_k|\bm{x}_k, \bm{R}_k)p(\bm{\Theta}_k|\bm{y}_{1:k-1}).
            \end{equation}
\end{itemize}

\begin{remark}
Due to the intractable integration and the nonlinearity involved in the predict-update cycle, the posterior PDF of joint latent variables has no closed-form analytical solution.
Most existing VB-based adaptive or robust Kalman filter methods~\cite{sarkka2009recursive, Huang2017TAC, ardeshiri2015approximate, ma2018multiple, zhu2022sliding, yu2023robust, xu2020adaptive, xia2021fine, zhu2021novel, zhu2021adaptive} focus on the linear state space models, whereas the posterior PDFs of state and noise covariance are updated via the coordinate ascent algorithm with necessitated conjugate prior PDF. 
The coordinate ascent algorithm is in general intractable for nonlinear or non-conjugate models. One solution is to approximate the intractable nonlinear integrals using conventional techniques, such as spherical cubature integration approximation~\cite{2017DongVariational}, and then update the variational hyperparameters.
The resulting optimization of this approximation is no longer substantiated to maximize the ELBO objective~\cite{gultekin2017nonlinear}. 
An alternative way is to optimize the ELBO objective through stochastic-gradient descent directly, which exhibits the slow convergence~\cite{10247583}. 
\end{remark}

\section{Adaptive Kalman Filtering using CVI method}
\subsection{Conjugate-computation Variational Inference}
As aforementioned, it is intractable to calculate posterior PDF $p(\bm{\Theta}_k|\bm{y}_{k})$ in \eqref{eq::prediction}-\eqref{eq::update}.
VB translates the intractable posterior PDF inference problem into finding an optimal variational distribution $q(\bm{\Theta}_k;\bm{\lambda}_k)$ that approximates the true posterior distribution $p(\bm{\Theta}_k|\bm{y}_k)$ with minimized Kullback-Leibler~(KL) divergence, where $\bm{\lambda}_k$ denotes the variational parameters. Minimizing the KL-divergence is equivalent to maximize the following ELBO $\mathcal{B}(\bm{\lambda}_k)$
\begin{equation}\label{eq::ELBO}
    q(\bm{\Theta}_{k}; \bm{\lambda}_k^*) \!=\! \arg \max_{\bm{\lambda}_k}  \underbrace{\mathbb{E}_{q}\left[\log p(\bm{\Theta}_{k}, \bm{y}_{k}) \!-\! \log q(\bm{\Theta}_{k}; \bm{\lambda}_k) \right]}_{\mathcal{B}(\bm{\lambda}_k)}.
\end{equation}

Based on the coordinate ascent for conjugate exponential models, computation of the posterior PDF can be achieved by simply adding sufficient statistics of the likelihood to the natural parameter of the prior, which is referred to as \emph{conjugate computations}. Unfortunately, conjugate computations exhibit low efficiency when the joint PDF $p(\bm{\Theta}_{k}, \bm{y}_{k})$ contains non-conjugate terms~(i.e., nonlinear state space model) because of the expectation $\mathbb{E}_{q}[\log p(\bm{\Theta}_{k}, \bm{y}_{k})]$ in \eqref{eq::ELBO} lacking a closed-form solution. Natural-gradient descent exploits the Riemannian geometry of exponential-family approximations to enhance the convergence rate, compared with the stochastic-gradient descent. CVI expands the application of natural gradients to complicated non-conjugate models by utilizing a stochastic mirror-descent strategy in the expectation-parameter space. 
Consider the following preliminaries.

\textbf{Assumption 1~[minimality]:} The variational distribution $q(\bm{z}_k|\bm{\lambda}_k)$ is in the \emph{minimal}\footnote{Properly, we consider that the exponential family $q(\bm{z}_k|\bm{\lambda}_k)$ is minimal if there is no $\bm{\lambda}_k$ such that $\bm{\lambda}_k^{\top} \bm{\phi}(\bm{z}_k) = 0$.} exponential family with its density written as 
\begin{equation}
    q(\bm{z}_k|\bm{\lambda}_k) = g(\bm{z}_k)\exp{\left(\bm{\lambda}_k^{\top}\bm{\phi}(\bm{z}_k) - A(\bm{\lambda}_k) \right)},
\end{equation}
where $g(\bm{z}_k)$ denotes the base measure, $\bm{\phi}(\bm{z}_k)$ and $\bm{\lambda}_k$ are respectively defined as the sufficient statistics and natural parameters, and $A(\bm{\lambda}_k)$ is the log-partition function.

The minimal representation implies that there is a one-to-one mapping between the mean parameter $\bm{\mu}_k \triangleq \mathbb{E}_q[\bm{\phi}_k]$ and the natural parameters $\bm{\lambda}_k$. Therefore, the ELBO optimization $\mathcal{B}(\bm{\lambda}_k)$ in natural parameters space can be expressed over $\bm{\mu}_k \in \mathcal{M}$, where $\mathcal{M}$ is the set of valid mean parameters space. This new objective function is denoted by $\mathcal{\bar B}(\bm{\mu}_k) \triangleq \mathcal{B}(\bm{\lambda}_k)$.

\textbf{Assumption 2~[conjugacy]:} The joint distribution $p(\bm{z}_k, \bm{y}_k)$ in \eqref{eq::ELBO} is assumed to be partitioned into two joint distributions, i.e., $p(\bm{z}_k, \bm{y}_k) \propto p_{\text{nc}}(\bm{z}_k, \bm{y}_k) p_{\text{c}}(\bm{z}_k, \bm{y}_k)$, including the conjugate terms $p_{\text{c}}(\bm{z}_k, \bm{y}_k)$ and non-conjugate terms $p_{\text{nc}}(\bm{z}_k, \bm{y}_k)$. 

This assumption can always be satisfied, e.g., by setting $p_{\text{c}}(\bm{z}_k, \bm{y}_k) = p(\bm{z}_k, \bm{y}_k)$ and $ p_{\text{nc}}(\bm{z}_k, \bm{y}_k) = 1$ for linear state space model. By categorizing the joint distribution into the conjugate part and the non-conjugate part, one can perform conjugate computations on the conjugate part and stochastic gradient descent on the non-conjugate part, enabling an effective and scalable inference. 

\textbf{Stochastic gradient descent:} Considering the ELBO optimization problem in \eqref{eq::ELBO}, the conjugate computation is intractable due to the non-conjugate terms. One can resort to the stochastic gradient descent, which optimizes the natural parameters $\bm{\lambda}_k$ iteratively by computing the gradient of the ELBO objective function, that is, 
\begin{equation}
\label{SGD}
    \bm{\lambda}^{(r+1)}_{k} = \bm{\lambda}^{(r)}_{k} + \beta_k^{(r)}\nabla_{\bm{\lambda}} \mathcal{B}(\bm{\lambda}_{k}),
\end{equation}
where $r$ is the iteration number, $\beta_k^{(r)}$ is the learning rate or step size, $\nabla_{\bm{\lambda}} \mathcal{B}(\bm{\lambda}_{k}) \triangleq \partial \mathcal{B}(\bm{\lambda}_{k}) / \partial \bm{\lambda}_{k}$ is a stochastic gradient of the lower bound at $\bm{\lambda}_{k} = \bm{\lambda}^{(r)}_{k}$. 

\begin{remark}
Stochastic gradient descent may converge slowly because it relies on an estimate of the stochastic gradient.
Natural gradient descent, exploiting the Riemannian geometry of variational distribution, uses the inverse Fisher information matrix as a preconditioning matrix to speed up the convergence. The natural parameters update with natural gradient is 
   \begin{equation}
   \label{NGD}
    \bm{\lambda}^{(r+1)}_{k} = \bm{\lambda}^{(r)}_{k} + \beta_k^{(r)} \bm{\mathrm{F}}^{-1}(\bm{\lambda}_k)\nabla_{\bm{\lambda}} \mathcal{B}(\bm{\lambda}_{k}),
\end{equation}  
where $\bm{\mathrm{F}}^{-1}(\bm{\lambda}_k)\nabla_{\bm{\lambda}} \mathcal{B}(\bm{\lambda}_{k})$ is defined as the natural gradient. 
\end{remark}

%\begin{remark}
Apparently, the inverse Fisher information matrix $\bm{\mathrm{F}}^{-1}(\bm{\lambda}_k)$ in \eqref{NGD} needs to be computed and stored explicitly in every iteration, which increases the computation burden and might be infeasible when the variational parameters are high-dimensional.
%\end{remark}

\textbf{Stochastic mirror descent~\cite{khan2017conjugate}:} Instead of updating in natural-parameter space, stochastic mirror descent avoids the computation of the Fisher information matrix by updating the parameters in expectation-parameter space. It establishes the equivalence with the natural gradient descent. The natural parameters can be updated as
\begin{equation}
     \bm{\lambda}^{(r+1)}_{k} = \bm{\lambda}^{(r)}_{k} + \beta_k^{(r)}\nabla_{\bm{\mu}} \mathcal{\bar B}(\bm{\mu}_{k}),
\end{equation}
where 
\begin{equation}
    \nabla_{\bm{\mu}} \mathcal{\bar B}(\bm{\mu}_{k}) = \left[ \dfrac{\partial{\bm{\mu}_{k}}}{\partial{\bm{\lambda}_{k}}} \right]^{-1} \dfrac{\partial{\mathcal{B}(\bm{\lambda}_{k})}}{\partial{\bm{\lambda}_{k}}} = \bm{\mathrm{F}}^{-1}(\bm{\lambda}_k)\nabla_{\bm{\lambda}} \mathcal{B}(\bm{\lambda}_{k})
\end{equation}
denotes the gradients of the ELBO with respect to~(w.r.t.) the expectation parameters $\bm{\mu}_k$.

\subsection{Adaptive Kalman Filtering with CVI}
Following the framework of recursive Bayesian filtering, the adaptive Kalman filtering encompasses three essential phases: initialization, prediction, and update. 
The initialization phase involves the modeling of the prior PDF of  
latent variables.

\subsubsection{Initialization} Assume that the joint conditional distribution of $\bm{\Theta}_{k-1}$ given measurements $\bm{y}_{1:k-1}$ is approximated as a product of distributions as follows:
\begin{equation}\label{Initialization}
    \begin{split}
        p(\bm{\Theta}_{k-1}|\bm{y}_{1:k-1}) =& \textbf{N}(\bm{x}_{k-1}|\hat{\bm{x}}_{k-1|k-1}, \bm{P}_{k-1|k-1})\times  \\
        &\textbf{IW}(\bm{Q}_{k-1}|\hat{t}_{k-1|k-1}, \bm{T}_{k-1|k-1})\times  \\
        &\textbf{IW}(\bm{R}_{k-1}|\hat{v}_{k-1|k-1}, \bm{V}_{k-1|k-1}),
    \end{split}
\end{equation}
where the distribution $\textbf{N}(\bm{x}|\bm{m}, \bm{P})$ denotes the random vector $\bm{x}$ obeys Gaussian distribution with mean $\bm{m}$ and covariance $\bm{P}$, and the distribution $\textbf{IW}(\bm{S}|\nu,\bm{\Psi})$ signifies that $\bm{S}$ obeys an inverse Wishart distribution with scale matrix $\bm{\Psi} \in\mathbb{R} ^{p\times p}$ and degrees of freedom parameter $\nu$ satisfying $\nu > p +1$. Then, $\mathbb{E}[\bm{S}^{-1}] = (\nu - p -1) \bm{\Psi}^{-1}$. 

\subsubsection{Prediction} Given the previous PDF $p(\bm{\Theta}_{k-1}|\bm{y}_{1:k-1})$, the predicted PDF can be factorized as 
\begin{equation}
    \label{Predicted-PDF}
    \begin{split}
    p(\bm{\Theta}_{k}|\bm{y}_{1:k-1}) =& \textbf{N}(\bm{x}_{k}|\hat{\bm{x}}_{k|k-1}, \bm{P}_{k|k-1}) \\
    &\times \textbf{IW}(\bm{Q}_{k}|\hat{t}_{k|k-1}, \bm{T}_{k|k-1}) \\
    &\times \textbf{IW}(\bm{R}_{k}|\hat{v}_{k|k-1}, \bm{V}_{k|k-1}).
    \end{split}
\end{equation}

It is worth noting that $\bm{R}_k$ is assumed to be independent with  $\bm{x}_k$ and $\bm{Q}_k$, while $\bm{x}_k$ and $\bm{Q}_k$ are coupled due to the relationship between predicted state covariance $\bm{P}_{k|k-1}$ and $\bm{Q}_k$.
By the fact that the covariance of $\bm{x}_k$ is $\bm{P}_{k|k-1}$ rather than $\bm{Q}_k$, the predicted PDF $p(\bm{x}_k,\bm{Q}_k|\bm{y}_{1:k-1})$ is non-conjugate, making the inference of the prediction step in \eqref{eq::prediction} intractable.
In the vein of~\cite{Huang2017TAC}, we regard $\bm{P}_{k|k-1}$ as the latent variable instead of $\bm{Q}_k$ to make the conjugate distribution available. Then, the joint latent variables are redefined as $\bm{\bar \Theta}_k \triangleq \{\bm{x}_k, \bm{P}_{k|k-1}, \bm{R}_k\}$. The predicted PDFs can be reconstructed as
\begin{equation}
\label{eq::Predicted-PDF}
    \begin{split}
        p(\bm{\bar \Theta}_k|\bm{y}_{1:k-1}) = & \textbf{N}(\bm{x}_k|\hat{\bm{x}}_{k|k-1},\bm{P}_{k|k-1}) \\
        &\times \textbf{IW}(\bm{P}_{k|k-1}|\hat{u}_{k|k-1}, \bm{U}_{k|k-1}) \\
        &\times \textbf{IW}(\bm{R}_{k}|\hat{v}_{k|k-1}, \bm{V}_{k|k-1})
    \end{split}
\end{equation}
with the parameters of the predicted PDF given by~\cite{Huang2017TAC}
\begin{align}
\label{eq::x_k|k-1}
    \hat{\bm{x}}_{k|k-1} = \,& \mathbb{E}[\bm{f}_{k}(\bm{x}_{k-1})], \\ \nonumber
    \bm{P}_{k|k-1} = \,& \mathbb{E}[( \bm{f}_{k}(\bm{x}_{k-1}) - \hat{\bm{x}}_{k|k-1} )(\cdot)^{\top}] + \hat{\bm{Q}}_{k},\\
    \label{eq::u_k|k-1}
    \hat{u}_{k|k-1} =\,& n + \tau_{P} +1,\\ \nonumber
    \bm{U}_{k|k-1} = \,&\tau_{P} \bm{P}_{k|k-1}, \\
    \label{eq::v_k|k-1}
    \hat{v}_{k|k-1} =\,&  \rho_{R} ( \hat{v}_{k-1|k-1} - m - 1) + m + 1,    \\ \nonumber
    \bm{V}_{k|k-1} =\,& \rho_{R} \bm{V}_{k-1|k-1},
\end{align}
where $\hat{\bm{Q}}_{k}$ is the nominal process noise covariance, $\tau_{P}>0$ is a tuning parameter, and $\rho_{R} \in (0,1]$ is the forgetting factor of $\bm{R}_k$. $n$ and $m$ are the dimensions of state and measurement, respectively. The initial $\bm{R}_0$ is assumed to be inverse Wishart distribution, i.e., $p(\bm{R}_0) = \bm{\mathrm{IW}}(\bm{R}_0;\hat{v}_{0|0},\bm{U}_{0|0})$.
The expectation of $\bm{R}_0$ satisfies that $\hat{\bm{R}}_0 = \bm{V}_{0|0}/(\hat{v}_{0|0}-m-1)$ with $\hat{v}_{0|0} = \tau_R + m + 1$ and $\bm{V}_{0|0} = \tau_R \hat{\bm{R}}_0$.

\subsubsection{Update} By the assumption that the likelihood function is a Gaussian distribution, the predicted PDF $p(\bm{\bar \Theta}_k|\bm{y}_{1:k-1})$ is updated with measurement $\bm{y}_k$ by
\begin{equation}
\label{eq::update-PDF}
\begin{split}
     p(\bm{\bar \Theta}_k|\bm{y}_{1:k}) \!\propto & \textbf{N}(\bm{y}_{k}|\bm{h}_k(\bm{x}_{k}), \bm{R}_{k}) \textbf{N}(\bm{x}_{k}|\hat{\bm{x}}_{k|k-1}, \bm{P}_{k|k-1})  \\
      &\times \textbf{IW}(\bm{P}_{k|k-1}|\hat{u}_{k|k-1}, \bm{U}_{k|k-1}) \\
      &\times \textbf{IW}(\bm{R}_{k}|\hat{v}_{k|k-1}, \bm{V}_{k|k-1}).
\end{split}
\end{equation}

It follows that the state $\bm{x}_k$ and measurement noise covariance $\bm{R}_k$ are coupled due to the likelihood function $p(\bm{y}_k|\bm{h}_k(\bm{x}_k), \bm{R}_k)$, 
for which the posterior PDF does not exist in closed forms. 
Nonetheless, we can approximate the posterior PDF via the CVI methods~\cite{khan2017conjugate}. Based on the \emph{mean-field} VB approximation where the latent variables are independent of each other and controlled by the corresponding hyperparameters, the variational distribution can be factorized as 
\begin{equation}
     q(\bm{\bar \Theta}_k;\bm{\lambda}_k) = q(\bm{x}_k;\bm{\lambda}_k^{\bm{x}}) 
     q(\bm{P}_{k|k-1};\bm{\lambda}_k^{\bm{P}}) q(\bm{R}_k;\bm{\lambda}_k^{\bm{R}}),
 \end{equation}
where $\bm{\lambda}_k \triangleq \{\bm{\lambda}_k^{\bm{x}}, \bm{\lambda}_k^{\bm{P}}, \bm{\lambda}_k^{\bm{R}}\}$ denotes the natural parameters of the corresponding variational distribution. The variational distribution, e.g., $q(\bm{x}_k;\bm{\lambda}_k^{\bm{x}})$, is the approximated distribution of the intractable posterior distribution $p(\bm{x}_k|\bm{y}_{1:k})$. 
In this context, adaptive Kalman filtering is necessitated to solve the following ELBO optimization problem,
\begin{equation}
    \label{eq::ELBO-m}
     \bm{\lambda}_k^* = \arg \max_{\bm{\lambda}_k}  \underbrace{\mathbb{E}_{q}\left[\log p(\bm{\bar \Theta}_k, \bm{y}_{k}) - \log q(\bm{\bar \Theta}_k; \bm{\lambda}_k) \right]}_{\mathcal{B}(\bm{\lambda}_k)}.
\end{equation}

Unlike~\cite{2017DongVariational, Huang2017TAC} that carried out the optimization in \eqref{eq::ELBO-m} directly within the space of the hyperparameter or natural-parameter $\bm{\lambda}_k$, we perform the stochastic mirror descent in the expectation-parameter space $\bm{\mu}_k$, avoiding high complexity of computing the
Fisher information matrix. In this way, the alternative ELBO $\bar{\mathcal{B}}(\bm{\mu}_k)$ is given by 
\begin{align}
\label{eq::B(u)}
\begin{split}
\bar{\mathcal{B}}(\bm{\mu}_k)
    &=  \mathbb{E}[\log p(\bm{x}_k|\hat{\bm{x}}_{k|k-1},\bm{P}_{k|k-1})]\\
    &+\mathbb{E}[\log p(\bm{P}_{k|k-1}|\hat{u}_{k|k-1},\bm{U}_{k|k-1})] \\
    &+ \mathbb{E}[\log p(\bm{R}_k|\hat{v}_{k|k-1},\bm{V}_{k|k-1})]  \\
    &+ \mathbb{E}[\log p(\bm{y}_k | \bm{h}_k(\bm{x}_k), \bm{R}_k)]- \mathbb{E}[\log q(\bm{x}_k; \bm{\mu}_k^{\bm{x}})] \\
    &- \mathbb{E}[\log q(\bm{P}_{k|k-1};\bm{\mu}_k^{\bm{P}})] - \mathbb{E}[\log q(\bm{R}_k; \bm{\mu}_k^{\bm{R}})].
\end{split}
\end{align}
The Gaussian distribution $\bm{x}_k$ and Inverse Wishart distribution $\bm{P}_{k|k-1}$, $\bm{R}_k$ belong to the minimal exponential family distribution, which is easily expressed as a function of its sufficient statistics $\bm{\phi}(\bm{z})$.
For state $\bm{x}_k \sim q(\bm{x}_k| \bm{m}_k, \bm{P}_k)$, the natural parameters $\bm{\lambda}_k^{\bm{x}}$ and expectation parameters $\bm{\mu}_k^{\bm{x}}$ are given by
    \begin{equation}
        \bm{\lambda}_k^{\bm{x}} = \left[ \begin{matrix}
           \bm{P}_k^{-1}\bm{m}_k \\
           - 0.5\bm{P}_k^{-1}
       \end{matrix}
       \right], 
       \bm{\mu}_k^{\bm{x}}
       = \left[\begin{matrix}
            \mathbb{E}_{q}(\bm{x}_k) \\
            \mathbb{E}_{q} (\bm{x}_k \bm{x}_k^{\top})
        \end{matrix}\right].
\end{equation}
For predicted state covariance $\bm{P}_{k|k-1} \sim \textbf{IW}(\bm{P}_{k|k-1}|u_k, \bm{U}_k)$ and measurement noise covariance $\bm{R}_{k} \sim \textbf{IW}(\bm{R}_{k}|v_k, \bm{V}_k)$,  the natural parameters $\bm{\lambda}_k^{\bm{P}}, \bm{\lambda}_k^{\bm{R}}$ and expectation parameters $\bm{\mu}_k^{\bm{P}}, \bm{\mu}_k^{\bm{R}}$ are given by
\begin{equation}
    \bm{\lambda}_k^{\bm{P}} = \left[ \begin{matrix}
        -0.5(u_k + n + 1) \\
        -0.5\bm{U}_k
    \end{matrix}
    \right], 
    \bm{\mu}_k^{\bm{P}} = \left[\begin{matrix}
            \mathbb{E}(\log \left| \bm{P}_{k|k-1} \right|)\\
            \mathbb{E}(\bm{P}_{k|k-1}^{-1})
        \end{matrix}\right],
\end{equation}
and 
\begin{equation}
    \bm{\lambda}_k^{\bm{R}} = \left[ \begin{matrix}
        -0.5(v_k + m + 1) \\
        -0.5\bm{V}_k
    \end{matrix}
    \right], 
    \bm{\mu}_k^{\bm{R}} = \left[\begin{matrix}
            \mathbb{E}(\log \left| \bm{R}_{k} \right| )\\
            \mathbb{E}(\bm{R}_{k}^{-1})
        \end{matrix}\right].
\end{equation}

Based on the conjugate-computation variational inference, we hereby utilize the adaptive Kalman filtering, referred to as the CVIAKF, to 
solve the ELBO optimization problem as follows
\begin{equation}
\label{optimization_B(u)}
     \bm{\lambda}_k^* = \arg \max_{\bm{\lambda}_k} \bar{\mathcal{B}}(\bm{\mu}_k).
\end{equation}

\textbf{Theorem 1.} Consider the adaptive Kalman filter for linear state space model with $\bm{f}_k(\bm{x}_{k}) = \bm{F}_k\bm{x}_k$ and $\bm{h}_k(\bm{x}_k)=\bm{H}_k\bm{x}_k$, where $\bm{F}_k$ and $\bm{H}_k$ are known state transition matrix and measurement matrix, respectively. The optimal natural parameters $\bm{\lambda}_k$ are updated as follows:
\begin{enumerate}
    \item Update the natural parameters $\bm{\lambda}_k^{\bm{x}}$ as
    \begin{equation}
        \label{x in linear case}
         \begin{split}
          \bm{\lambda}_k^{\bm{x}}(1) =& \mathbb{E}[\bm{P}_{k|k-1}^{-1}] \hat{\bm{x}}_{k|k-1} + \bm{H}_k^{\top}\mathbb{E}[\bm{R}_k^{-1}]\bm{y}_k,
            \\
          \bm{\lambda}_k^{\bm{x}}(2) =&  -0.5\mathbb{E}[\bm{P}_{k|k-1}^{-1}] -0.5 \bm{H}_k^{\top}\mathbb{E}[\bm{R}_k^{-1}]\bm{H}_k,
        \end{split}
    \end{equation}
    with 
    \begin{equation*}
    \begin{split}
        \mathbb{E}[\bm{P}_{k|k-1}^{-1}] &= (\hat{u}_{k|k} - n -1) \bm{U}_{k|k}^{-1}, \\
        \mathbb{E}[\bm{R}_k^{-1}] &= (\hat{v}_{k|k} - m -1) \bm{V}_{k|k}^{-1}.
    \end{split}
    \end{equation*}   
    \item Update the natural parameters $\bm{\lambda}_k^{\bm{P}}$ as
    \begin{equation}
    \label{Q in linear case}
    \begin{split} 
    \bm{\lambda}_k^{\bm{P}}(1) =& -0.5(\hat{u}_{k|k-1} + n + 2),
        \\
     \bm{\lambda}_k^{\bm{P}}(2) =& -0.5(\bm{U}_{k|k-1} + \bm{C}_k),
    \end{split}
    \end{equation}
    with $\bm{C}_k = (\hat{\bm{x}}_{k|k} - \hat{\bm{x}}_{k|k-1})(\cdot)^{\top} + \bm{P}_{k|k}$.
    \item Update the natural parameters $\bm{\lambda}_k^{\bm{R}}$ as
    \begin{equation}
    \label{R in linear case}
    \begin{split} 
    \bm{\lambda}_k^{\bm{R}}(1) &= -0.5(\hat{v}_{k|k-1} + m + 2),           \\
    \bm{\lambda}_k^{\bm{R}}(2) &= -0.5(\bm{V}_{k|k-1} + \bm{A}_k),
    \end{split}
\end{equation}
with $\bm{A}_k = (\bm{y}_k - \bm{H}_k\hat{\bm{x}}_{k|k})(\cdot)^{\top} + \bm{H}_k\bm{P}_{k|k}\bm{H}_k^{\top}$.
\end{enumerate}
\textbf{Proof. } See Appendix A. $\hfill \Large\square$

\textbf{Proposition 1.} For linear models, CVIAKF is equivalent to VBAKF proposed in \cite{Huang2017TAC}. Also, CVIAKF can be transformed into an information filter when $\bm{Q}_k$ and $\bm{R}_k$ are known.

\textbf{Proof. } See Appendix B. $\hfill \Large\square$

\textbf{Theorem 2.} \label{CVIAKF for nonlinear}Considering the adaptive Kalman filter for nonlinear state space model \eqref{stochastic-systems}, the optimal natural parameters $\bm{\lambda}_k$ can be iteratively updated as follows:

(1) Update the natural parameters $\bm{\lambda}_k^{\bm{x}}$ iteratively as
\begin{equation}
\label{state in nonlinear case}
\begin{split}
   (\bm{P}_{k|k}^{-1})^{(r+1)} = & 0.5(\beta_k^{(r)})^2 \hat{\bm{\mathrm{G}}}_k \bm{P}_{k|k}^{(r)} \hat{\bm{\mathrm{G}}}_k + (1 - \beta_k^{(r)}) (\bm{P}_{k|k}^{-1})^{(r)} \\
   &+ \beta_k^{(r)} (\mathbb{E}[\bm{P}_{k|k-1}^{-1}] + \nabla_{\bm{P}_{k|k}} \mathcal{D}(\bm{\mu}_{\bm{x}}) ) \\
    \hat{\bm{x}}_{k|k}^{(r+1)} = &\hat{\bm{x}}_{k|k}^{(r)} -0.5 \beta_k^{(r)}  \bm{P}_{k|k} ^{(r+1)} \nabla_{\hat{\bm{x}}_{k|k}} \mathcal{D}(\bm{\mu}_{\bm{x}}) \\
    &+ \beta_k^{(r)}  \bm{P}_{k|k} ^{(r+1)} \mathbb{E}[\bm{P}_{k|k-1}^{-1}](\hat{\bm{x}}_{k|k-1} - \hat{\bm{x}}_{k|k}^{(r)}) 
    \end{split}
\end{equation}
with
\begin{equation*}
   \label{eq::gradients}
    \begin{split}
    \nabla_{\hat{\bm{x}}_{k|k}} \mathcal{D}(\bm{\mu}_{\bm{x}}) = & -\frac{2}{S}\sum_{s = 1} ^{S} \bm{H}_k (\bm{g}(\bm{\epsilon}_k^s))^{\top}\mathbb{E}[\bm{R}_k^{-1}] \\
    &\times (\bm{y}_k - \bm{h}_k(\bm{g}(\bm{\epsilon}_k^s))), \\
    \nabla_{\bm{P}_{k|k}} \mathcal{D}(\bm{\mu}_{\bm{x}}) 
         = & -\frac{1}{S}\sum_{s = 1} ^{S} \bm{H}_k(\bm{g}(\bm{\epsilon}_k^s))^{\top}\mathbb{E}[\bm{R}_k^{-1}]\\
         &\times (\bm{y}_k - \bm{h}_k(\bm{g}(\bm{\epsilon}_k^s)))
         (\bm{\epsilon}_k^{s})^{\top} \bm{L}_k^{-1},
        \\
          \hat{\bm{\mathrm{G}}}_k = &(\bm{P}_{k|k}^{-1})^{(r)} - \mathbb{E}[\bm{P}_{k|k-1}^{-1}] - \nabla_{\bm{P}_{k|k}} \mathcal{D}(\bm{\mu}_{\bm{x}}), 
    \end{split}
\end{equation*}
where $\bm{g}(\bm{\epsilon}_k^s) = \hat{\bm{x}}_{k|k} + \bm{L}_k\bm{\epsilon}_k^s$ with $\bm{L}_k \bm{L}_k^{\top} = \bm{P}_{k|k}$, $\bm{\epsilon}^{s}_k \overset{\mathrm{iid}}{\sim} \bm{\mathrm{N}}(\bm{\mathrm{0}}, \bm{\mathrm{I}}_n)$, $s = 1,2,\cdots,S$ with $S$ being the total sampling number. $\bm{H}_k(\bm{g}(\bm{\epsilon}^s))$ denotes the Jacobian matrix of $\bm{h}_k(\bm{g}(\bm{\epsilon}^s))$. $\beta_k^{(r)} \in (0, 1]$ is the learning rate, $r$ is the iteration number.

(2) The updates of natural parameters $\bm{\lambda}_{k}^{\bm{P}}$ and $\bm{\lambda}_k^{\bm{R}}$ have the same forms as \eqref{Q in linear case} and \eqref{R in linear case}, where $\bm{A}_k$ is corrected as
    \begin{equation}
    \label{eq::A in nonlinear case}
         \bm{A}_k = \sum_{s = 1}^{S} (\bm{y}_k - \bm{h}_k(\hat{\bm{x}}_{k|k} + \bm{L}_k\bm{\epsilon}_k^{s}))(\cdot)^{\top}.
    \end{equation}

 \textbf{Proof. } See Appendix C. $\hfill \Large\square$

\begin{remark}
The update of natural parameters $\bm{\lambda}_k^{\bm{x}}$ can be split into a conjugate part and a non-conjugate part. For the non-conjugate part arising from the nonlinear measurement function $\bm{h}_k(\bm{x}_k)$, the stochastic mirror gradient method needs to compute the intractable gradients w.r.t. the expectation parameters. To address this problem, the gradients w.r.t. the expectation parameters are transformed into gradients w.r.t. variational hyperparameters via the chain rule, where the reparameterization trick is employed to reduce the variance of gradient approximation. Meanwhile, the original stochastic mirror gradient method may suffer from numerical instability issues. The line-search approach~\cite{lin2020handling} is exploited to ensure the updated state covariance matrix $\bm{P}_{k|k}$ is always positive-definite. 
\end{remark}
 
\subsection{Summary and Discussion}
We have proposed the CVIAKF method for joint state estimation and noise parameter identification. For linear systems, CVIAKF has a closed-loop analytical solution and is equivalent to the information forms of VBAKF method. For nonlinear systems, CVIAKF updates the natural parameters of joint latent variables via the stochastic mirror gradient method incorporating the reparameterization trick and positive definite constraint. The proposed CVIAKF method can be summarized in \textbf{Algorithm}\ref{alg::CVIAKF}.

\begin{algorithm}[t]
\caption{\textbf{CVIAKF}~(at time $k$)}
\label{alg::CVIAKF}
    \KwInput {$\bm{y}_k$, $q(\bm{x}_{k-1})$,  $q(\bm{P}_{k-1})$ and $q(\bm{R}_{k-1})$;}
    \KwOutput {$q(\bm{x}_{k})$, $q(\bm{P}_{k|k-1})$ and $q(\bm{R}_{k})$;}

    \underline{\textbf{Time Prediction:}} compute the predicted PDFs

    $p(\bm{x}_k|\bm{y}_{1:k-1})$ via \eqref{eq::x_k|k-1};

    $p(\bm{P}_{k|k-1}|\bm{y}_{1:k-1})$ via \eqref{eq::u_k|k-1};

    $p(\bm{R}_k|\bm{y}_{1:k-1})$ via \eqref{eq::v_k|k-1};

    \underline{\textbf{Measurement Update:}} \tcp*{Iteratively update the variational PDFs}
    
    Set $q^{(0)}(\bm{x}_k) = p(\bm{x}_k|\bm{y}_{1:k-1})$,  $q^{(0)}(\bm{P}_{k|k-1}) = p(\bm{P}_{k|k-1}|\bm{y}_{1:k-1})$, $q^{(0)}(\bm{R}_k) = p(\bm{R}_k|\bm{y}_{1:k-1})$;

    \While{not convergence}  
    {
      \tcc{CVIAKF for linear models~(Theorem 1)}
        \If {$\bm{f}_k(\bm{x}_k)$ and $\bm{h}_k(\bm{x}_k)$ are linear} 
        {
            Update posterior $q(\bm{x}_k|\bm{\lambda}_k^{\bm{x}})$ via \eqref{x in linear case};
            
            Update posterior $q(\bm{P}_{k|k-1}|\bm{\lambda}_k^{\bm{P}})$ via \eqref{Q in linear case}; 

            Update posterior $q(\bm{R}_k|\bm{\lambda}_k^{\bm{R}})$ via \eqref{R in linear case};
  
        }
        \tcc{CVIAKF for nonlinear models~(Theorem 2)}
        \Else
        {
            Update posterior $q(\bm{x}_{k}|\bm{\lambda}_k^{\bm{x}})$ via \eqref{state in nonlinear case}  \tcp*{stochastic mirror gradient descent}

            Update posterior $q(\bm{P}_{k|k-1}|\bm{\lambda}_k^{\bm{P}})$ via \eqref{Q in linear case}; 

            Update posterior $q(\bm{R}_k|\bm{\lambda}_k^{\bm{R}})$ via \eqref{R in linear case} with $\bm{A}_k$ given by \eqref{eq::A in nonlinear case}.
        }
    }
\end{algorithm}

\section{Results For Synthetic Data}
\subsection{Simulation Configuration}
The simulation encompasses four distinct scenarios for tracking maneuvering targets, outlined as follows:
\begin{itemize}
    \item \textbf{S1}: Adaptive linear filtering with time-varying periodic noise covariances~\cite{ardeshiri2015approximate, Huang2017TAC};
    \item \textbf{S2}: Adaptive linear filtering with time-varying piecewise noise covariances~\cite{sarkka2009recursive, 2017DongVariational};
    \item \textbf{S3}: Adaptive nonlinear filtering with time-varying periodic noise covariances;
    \item \textbf{S4}: Adaptive nonlinear filtering with time-varying piecewise noise covariances.
\end{itemize}

The kinematic state of a target, denoted as $\bm{x}_k = [x_k, \dot x_k, y_k, \dot y_k]^{\top}$, comprises the target's position $[x_k, y_k]^{\top}$ and velocity $[\dot x_k, \dot y_k]^{\top}$. 
The initial state $\bm{x}_0 = [5e^5\text{m}, -100\text{m/s}, 5e^5\text{m}, -100\text{m/s}]^{\top}$, the total simulated steps $N=300$, and sampling period $T=1\text{s}$. In the simulations, the initial states $ \bm{\hat x}_{0|0}$ for filters are produced randomly from $\textbf{N}(\bm{x}_0, \bm{P}_0)$ in each turn with $\bm{P}_0 = \text{diag}([100\text{m}^2, 1\text{m}^2/\text{s}^2, 100\text{m}^2, 1\text{m}^2/\text{s}^2])$ for linear systems and $\bm{P}_0 = \text{diag}([100^2\text{m}^2, 10^2\text{m}^2/\text{s}^2, 100^2\text{m}^2, 10^2\text{m}^2/\text{s}^2])$ for nonlinear systems.
The state transition matrix $\bm{F}_k$ and nominal process noise covariance $\bm{Q}_0$ are given by 
\begin{equation*}
\bm{F}_k =  \textbf{I}_2 \otimes \begin{bmatrix}
1 & T \\
0 & 1 \\
\end{bmatrix}, \quad 
   \bm{Q}_0 =  \textbf{I}_2 \otimes \begin{bmatrix}
T^3/3 & T^2/2 \\
T^2/2 & T \\
\end{bmatrix},
\end{equation*}
where $\textbf{I}_2$ is $2 \times 2$ identity matrix. For linear scenarios \textbf{S1} and \textbf{S2}, the measurements $\bm{y}_k = [x_k, y_k]^T$, the measurement matrix $\bm{h}^c_k$ and nominal measurement noise covariance $\bm{R}^c_0$ are
\begin{equation*}
\bm{h}^c_0 = \begin{bmatrix}
    1 & 0 & 0 & 0 \\
    0 & 0 & 1 &0
\end{bmatrix}, 
\bm{R}^c_0 = \begin{bmatrix}
    10^4\text{m}^2 & 100\text{m}^2 \\
    100\text{m}^2 & 10^4\text{m}^2
\end{bmatrix}.
\end{equation*}
For nonlinear scenarios \textbf{S3} and \textbf{S4},  $\bm{y}_k = [r_k, a_k]^T$ consists of range $r_k$ and azimuth $a_k$. 
The measurement function $\bm{h}^{nc}_k$ and nominal measurement noise covariance $\bm{R}^{nc}_0$ are
\begin{equation*}
\bm{h}^{nc}_k = \begin{bmatrix}
     \sqrt{x_k^2 + y_k^2} \\
   \arctan\left({y_k}/{x_k}\right)
\end{bmatrix}, 
\bm{R}^{nc}_0 = \begin{bmatrix}
    100\text{m}^2 & 0\text{m$\cdot$rad} \\
    0\text{m$\cdot$rad} & 10^{-6}\text{rad}^2
\end{bmatrix}.
\end{equation*}

The process noise covariance $\bm{Q}_k$ and measurement noise covariance $\bm{R}_k$ for different scenarios are given in TABLE.~\ref{Tab::Scenario}
\begin{table}[h]%调节图片位置，h：浮动；t：顶部；b:底部；p：当前位置
    \centering
    \caption{Scenario parameters}
    \renewcommand{\arraystretch}{1}
    \label{Tab::Scenario}
    \begin{tabular}{c|c|c}
    \hline \hline
        \textbf{Scenario} & $\bm{Q}_k$ & $\bm{R}_k$ \\ \hline
        \textbf{S1} & $\left(10 + 5 \cos{(\pi k/N)} \right)\bm{Q}_0$ 
                    & $\left(1 + 0.5\cos{(\pi k/N)} \right)\bm{R}_0^c$ \\ \hline
        \textbf{S2} & $\begin{cases}
                        \bm{Q}_0, & k \in [1, 100) \\
                        5\bm{Q}_0, & k \in [100, 200) \\
                        \bm{Q}_0, & k \in [200,300]
                    \end{cases}$ 
                    & $\begin{cases}
                        \bm{R}_0^c, & k \in [1, 100) \\
                        \bm{R}_0^c, & k \in [100, 200) \\
                        5\bm{R}_0^c, & k \in [200, 300]
                    \end{cases}$  \\ \hline
        \textbf{S3} & $\left(100 + 50 \cos{(\pi k/N)} \right)\bm{Q}_0$ 
                    & $\left(1 + 0.5\cos{(\pi k/N)} \right)\bm{R}_0^{nc}$ \\ \hline
        \textbf{S4} & $\begin{cases}
                        \bm{Q}_0, & k \in [1, 100) \\
                        100\bm{Q}_0, & k \in [100, 200) \\
                        \bm{Q}_0, & k \in [200, 300]
                    \end{cases}$ 
                    & $\left(1 + 0.5\cos{(\pi k/N)} \right)\bm{R}_0^{nc}$  \\ \hline
    \end{tabular}
\end{table}

The following adaptive Kalman filtering methods are compared with the proposed CVIAKF:
\begin{itemize}
    \item KFTCM: KF with true $\bm{Q}_k$ and $\bm{R}_k$;
    \item KFNCM: KF with nominal $\bm{Q}_0$ and $\bm{R}^c_0$;
    \item VBAKF~\cite{Huang2017TAC}: linear AKF with unknown $\bm{Q}_k, \bm{R}_k$; 
    \item VBACIF~\cite{2017DongVariational}: nonlinear AKF with unknown $\bm{R}_k$.
    \item SSVBAKF~\cite{10247583}: nonlinear AKF with unknown $\bm{Q}_k$.
    \item VBAKF-CM: VBAKF for nonlinear systems combined with unbiased measurement conversion~\cite{6850165}.
\end{itemize}

For linear cases \textbf{S1} and \textbf{S2}, we compare CVIAKF with KFTCM, KFNCM, and VBAKF. 
VBACIF, SSVBAKF, and VBAKF-CM are compared with CVIAKF for nonlinear cases \textbf{S3} and \textbf{S4}. The root mean square error~(RMSE) and average RMSE~(ARMSE) over time are used to evaluate the estimation performance of different methods
\begin{equation}
    \text{RMSE}(k) \triangleq \sqrt{{\dfrac{1}{M}\sum_{i=1}^M (\hat{x}_k^i - x_k^i)^2 + (\hat{y}_k^i - y_k^i)^2}},
\end{equation}
where $(\hat x_k^i, \hat y_k^i)$ and $(x_k^i, y_k^i)$ are the estimated and true values at the $i-$th Monte Carlo run with $M$ being the number of Monte Carlo runs. 

The parameters for different scenarios are set as follows:
\begin{itemize}
    \item For linear cases \textbf{S1} and \textbf{S2}, $\hat{\bm{Q}}_0 = \alpha \bm{\mathrm{I}}_4$, $\hat{\bm{R}}_0 = \beta \bm{\mathrm{I}}_2$ with $\alpha = 10$ and $\beta = 100$, $\tau_{R} = \tau_{P} = 3$, $\rho_{R} = 1 - \exp(-4)$, iteration threshold $\delta = 1e-7$.
    \item For nonlinear cases \textbf{S3} and \textbf{S4}, $\tau_{P} =3$, $\tau_R = 6$,  $\rho_{R} = 1 - \exp(-4)$, $\hat{\bm{Q}}_0 = \alpha \bm{\mathrm{I}}_4$ with $\alpha = 20$, $\hat{\bm{R}}_0 = \bm{R}_0$, iteration threshold $\delta = 1e-7$. The learning rate $\beta = 0.23$, and the samples $S = 1000$. Let $\bm{Q}_k = \bm{Q}_0$ and $\bm{R}_k = \bm{R}_0$ for VBACIF and SSVBAKF, respectively. 
\end{itemize}

\begin{figure}[htbp]
    \centering
    \subfloat[RMSE in position]{\includegraphics[scale=0.45]{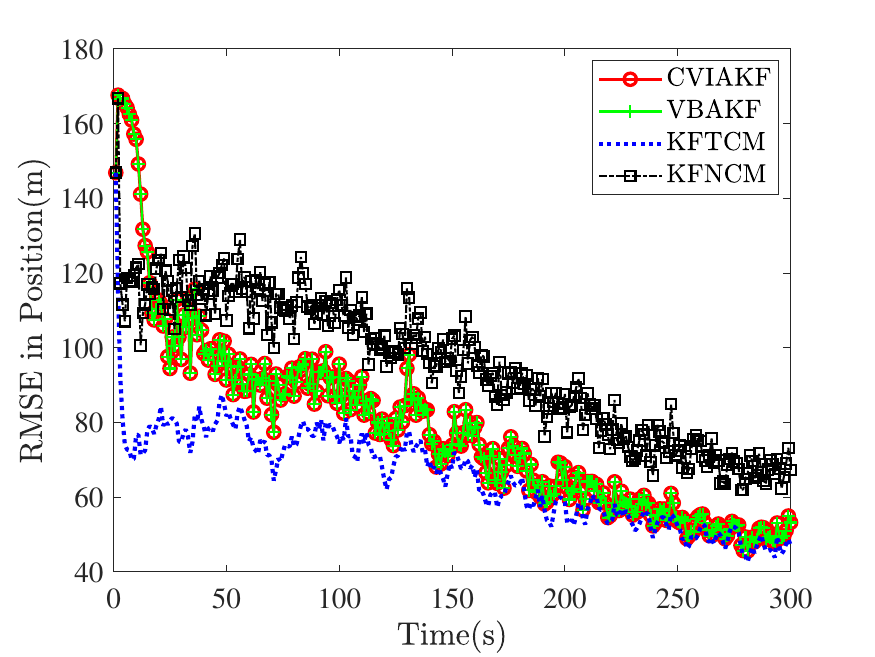}} \\
    \subfloat[RMSE in velocity]{\includegraphics[scale=0.45]{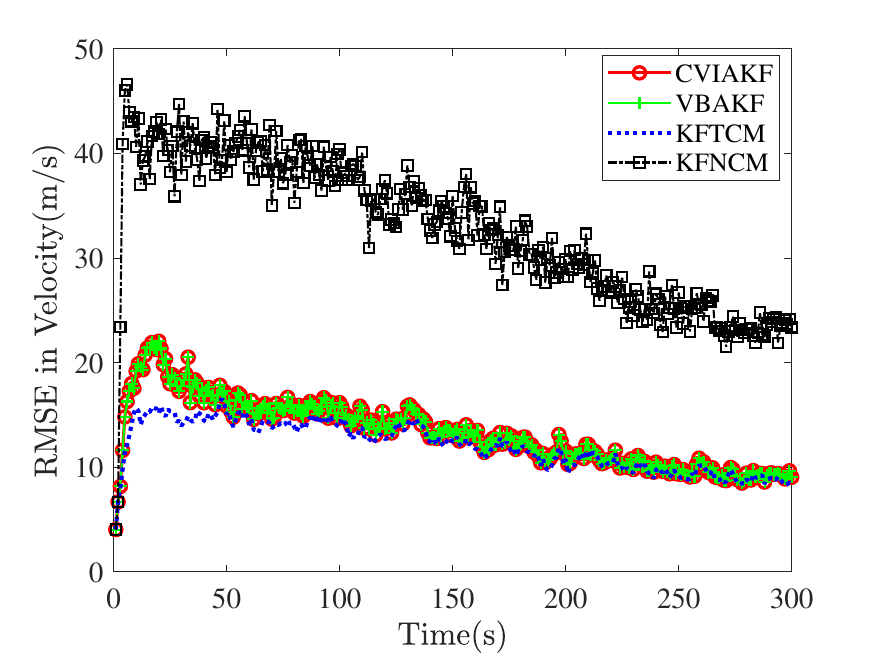}} 
    \caption{RMSE comparison for simulated scenario \textbf{S1}}
    \label{S1::RMSE}
\end{figure}

\begin{figure}[htbp]
    \centering
    \subfloat[RMSE in position]{\includegraphics[scale=0.45]{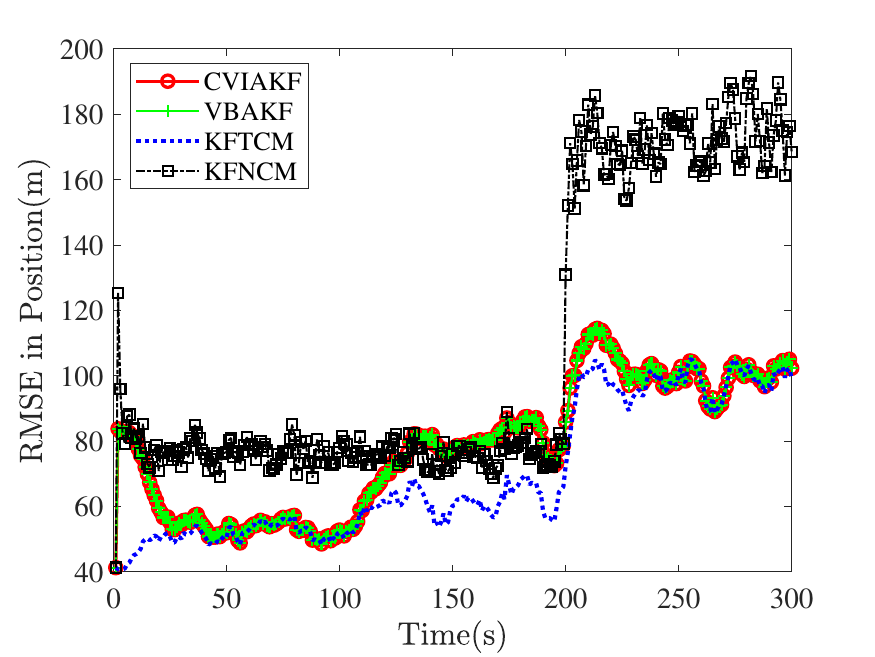}} \\
    \subfloat[RMSE in velocity]{\includegraphics[scale=0.45]{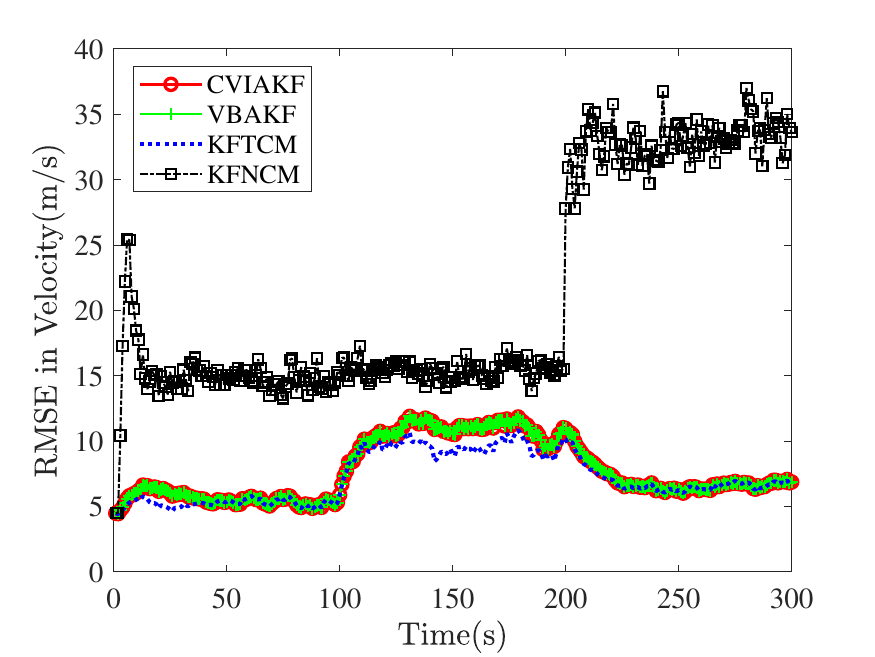}} 
    \caption{RMSE comparison for simulated scenario \textbf{S2}}
    \label{S2::RMSE}
\end{figure}

For each scenario, 100 Monte Carlo runs are accomplished on a computer equipped with a Legion Y9000P IRX8H CPU operating at 2.20 GHz. The comparison of RMSEs for scenario \textbf{S1} and \textbf{S2} are depicted in Fig.~\ref{S1::RMSE} and Fig.~\ref{S2::RMSE}, respectively. It is observed that CVIAKF has the same RMSEs as VBAKF, demonstrating their equivalence in the linear scenarios. 
CVIAKF outperforms KFNCM and is comparable to KFTCM as time increases. In particular, CVIAKF and VBAKF exhibit good estimation performance for scenario \textbf{S1} when the noise covariances $\bm{Q}_k$ and $\bm{R}_k$ are periodic time-varying. 
For scenario \textbf{S2} with time-varying piecewise noise covariances, the estimation performance of CVIAKF and VBAKF approach that of KFTCM when the noise covariance remains unchanged ($k \in [0, 100)$) or the measurement noise $\bm{R}_k$ changes but deteriorates when the process noise $\bm{Q}_k$ changes ($k \in [100, 200)$) abruptly. 
This is because the performance of CVIAKF and VBAKF is dependent on the initial values of $\bm{Q}_k$.
Abrupt changes in $\bm{Q}_k$ can lead to a deterioration of their performance~\cite{huang2020variational}.
Tables \ref{table::ARMSE for linear system} shows the ARMSEs of position and velocity, which indicates the effectiveness of CVIAKF. The computational cost of CVIAKF and VBAKF is about ten times higher than that of standard Kalman filtering. This is mainly because CVIAKF and VBAKF contain iterative loop processing among state estimation and noise covariance identification, and the average iterations are 9 and 7 for \textbf{S1} and \textbf{S2}, respectively.

\begin{table}[htbp]
    \centering
     \caption{ARMSE comparison for linear scenarios}
    \begin{tabular}{c|c|c|c|c}
    \hline \hline
     \textbf{Scenarios} & \multicolumn{2}{c|}{\textbf{S1}} & \multicolumn{2}{c}{\textbf{S2}} \\ \hline
     \textbf{Methods}  & Position & Velocity &  Position & Velocity \\ \hline
     \textbf{KFNCM}    & 95.21 & 32.79 & 108.54& 21.18 \\ \hline
     \textbf{VBAKF}    & 79.01 & 13.27 & 78.24 & 7.74 \\ \hline
     \textbf{CVIAKF}   & 79.01 & 13.27 & 78.24 & 7.74 \\ \hline
     \textbf{KFTCM}    & 65.34 & 12.07 & 69.87 & 7.19 \\ \hline
    \end{tabular}
    \label{table::ARMSE for linear system}
\end{table}

\begin{figure}[htbp]
    \centering
    \subfloat[RMSE in position]{\includegraphics[scale=0.45]{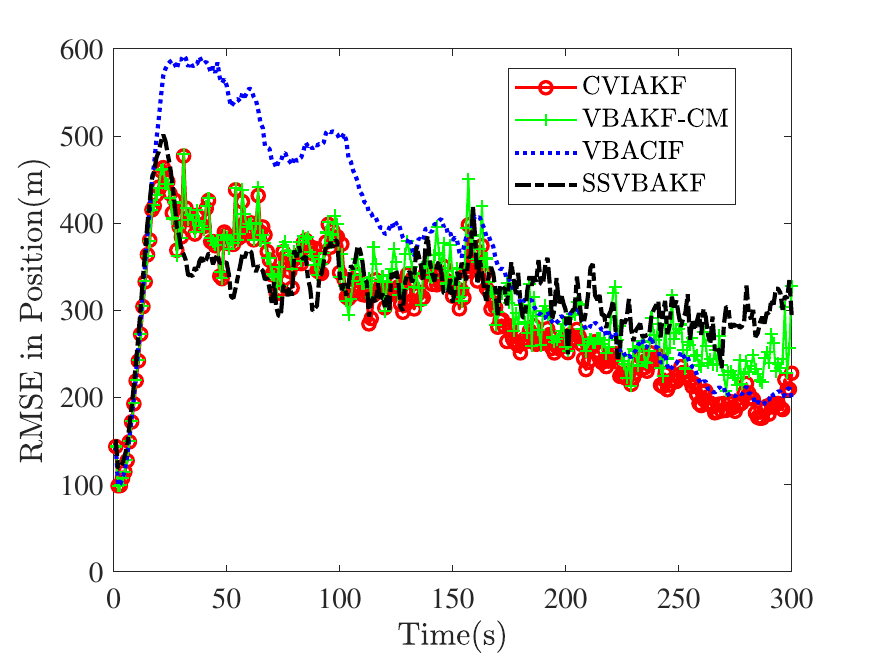}} \\
    \subfloat[RMSE in velocity]{\includegraphics[scale=0.45]{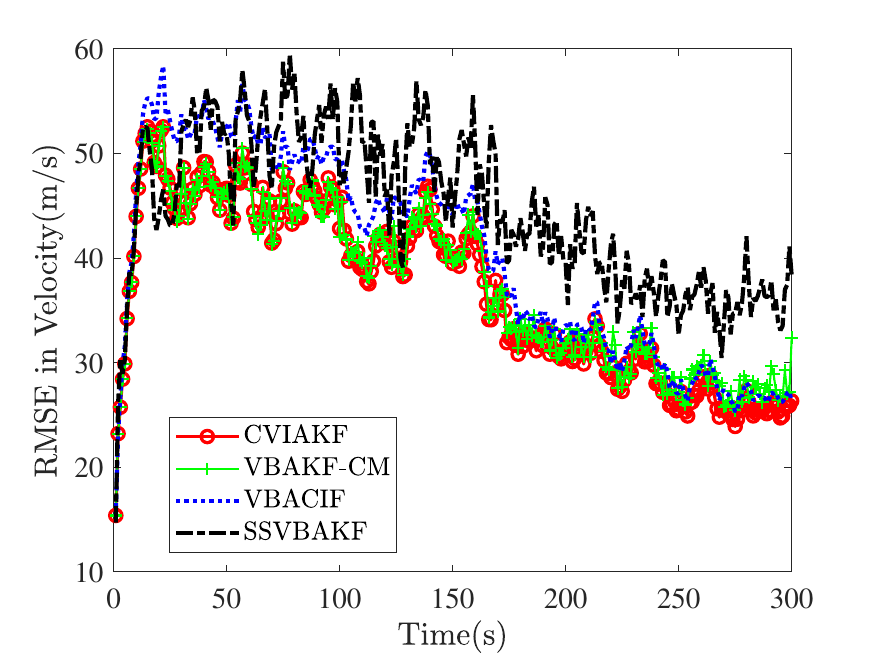}} 
    \caption{RMSE comparison for simulated scenario \textbf{S3}}
    \label{S3::RMSE}
\end{figure}

The RMSEs comparison for nonlinear scenarios \textbf{S3} and \textbf{S4} are shown in Fig.~\ref{S3::RMSE} and Fig.~\ref{S4::RMSE}, respectively. It is shown that CVIAKF outperforms existing VB-based adaptive filters. This is because both CVIAKF and VBAKF-CM estimate the unknown $\bm{Q}_k$ and $\bm{R}_k$, which is better than VBACIF estimating only $\bm{R}_k$ and SSVBAKF estimating only $\bm{Q}_k$. Meanwhile, CVIAKF directly optimizes the non-conjugate ELBO via the stochastic mirror gradient descent method. VBAKF-CM carries out the nonlinear adaptive filtering into two subproblems: linear approximation based on unbiased measurement conversion and linear adaptive filtering. In cases where the measurement noise covariance $\bm{R}_k$ is unknown, there may be significant errors in the conversion process. Table.\ref{table::ARMSE for nonlinear system} illustrates the ARMSE comparison for different methods, indicating that CVIAKF is the best, followed by VBAKF-CM and SSVBAKF, and VBACIF is the worst. All comparison methods are based on VB. The average iterations of VBACIF, VBAKF-CM, SSVBAKF and CVIAKF are 5, 9, 120, 27 for scenario \textbf{S3}, and 5, 8, 68, 24 for scenario \textbf{S4}, respectively. The fast convergence of VBACIF and VBAKF-CM can be attributed to their use of conjugate computations. On the other hand, SSVBAKF and CVIAKF require more iterations to update their hyperparameters since they rely on stochastic gradient optimization. However, CVIAKF uses stochastic mirror gradient descent to accelerate the convergence of stochastic gradient descent, which ultimately results in fewer iterations than SSVBAKF.

\begin{figure}[htbp]
    \centering
    \subfloat[RMSE in position]{\includegraphics[scale=0.45]{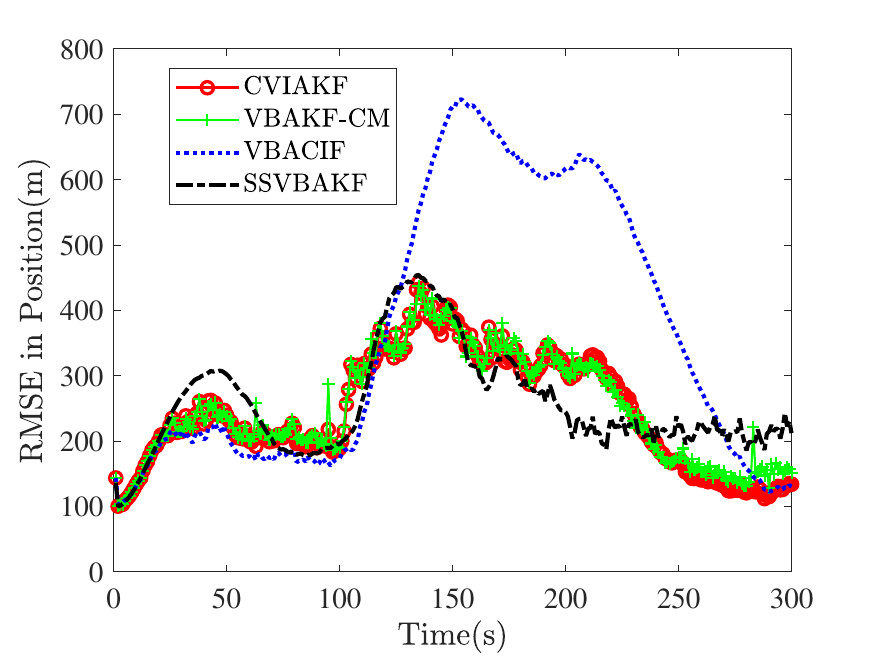}} \\
    \subfloat[RMSE in velocity]{\includegraphics[scale=0.45]{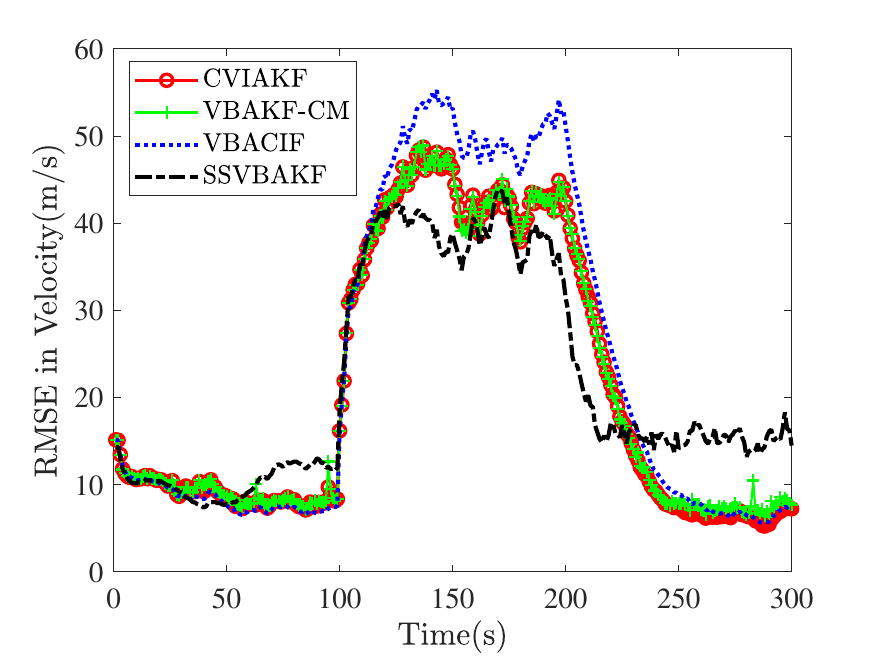}} 
    \caption{RMSE comparison for simulated scenario \textbf{S4}}
    \label{S4::RMSE}
\end{figure}

\begin{table}[htbp]
    \centering
     \caption{ARMSE comparison for nonlinear scenarios}
    \begin{tabular}{c|c|c|c|c}
    \hline \hline
     \textbf{Scenarios} & \multicolumn{2}{c|}{\textbf{S3}} & \multicolumn{2}{c}{\textbf{S4}} \\ \hline
     \textbf{Methods}  & Position & Velocity &  Position & Velocity \\ \hline
     \textbf{VBACIF}   & 363.55 & 40.46 & 362.55& 23.71 \\ \hline
     \textbf{SSVBAKF}  & 322.78 & 44.72 & 256.51 & 21.61 \\ \hline
     \textbf{VBAKF-CM} & 314.71 & 37.59 & 250.52 & 21.52 \\ \hline
     \textbf{CVIAKF}   & \textbf{\textcolor{blue}{295.19}} &  \textbf{\textcolor{blue}{37.03}} & \textbf{\textcolor{blue}{245.84}} & \textbf{\textcolor{blue}{21.32}} \\ \hline
    \end{tabular}
    \label{table::ARMSE for nonlinear system}
\end{table}

\section{Results For Real-World Data}
\begin{figure}[htbp]
    \centering
    \subfloat[\textbf{R1}]{\label{fig::R1} \includegraphics[scale=0.45]{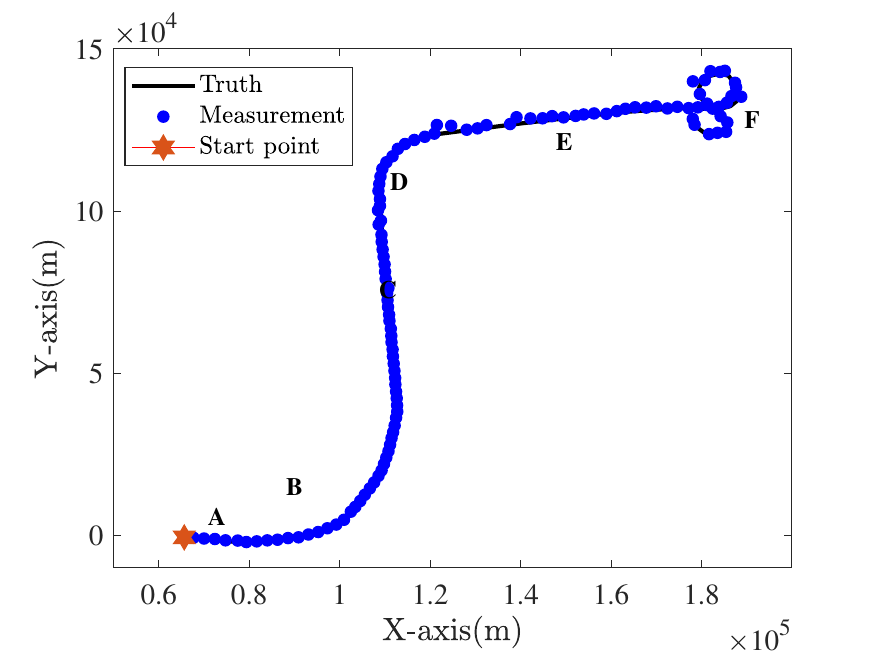}} \\
    \subfloat[\textbf{R2}]{\label{fig::R2} \includegraphics[scale=0.45]{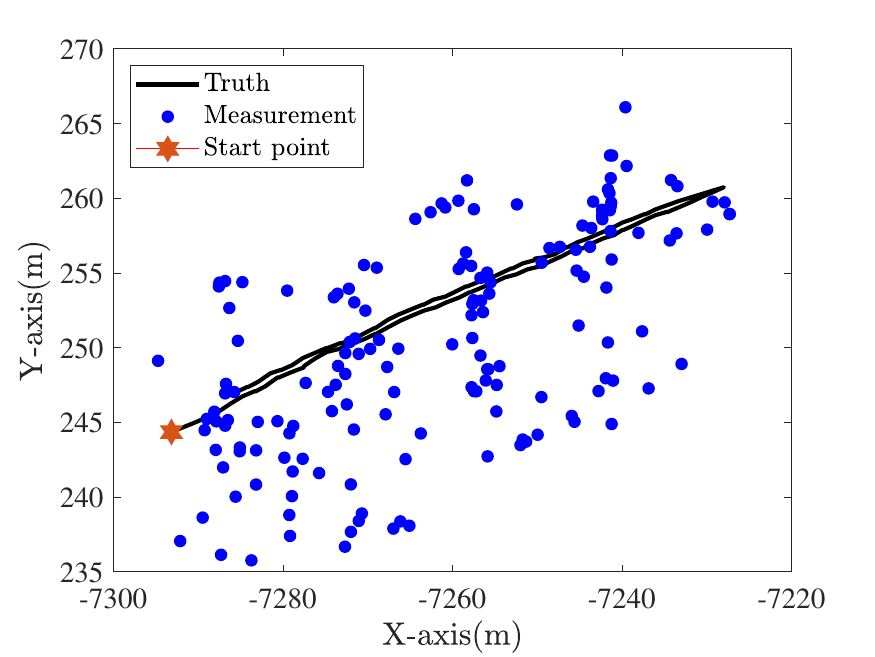}} \\
    \subfloat[\textbf{R3}]{\label{fig::R3} \includegraphics[scale=0.45]{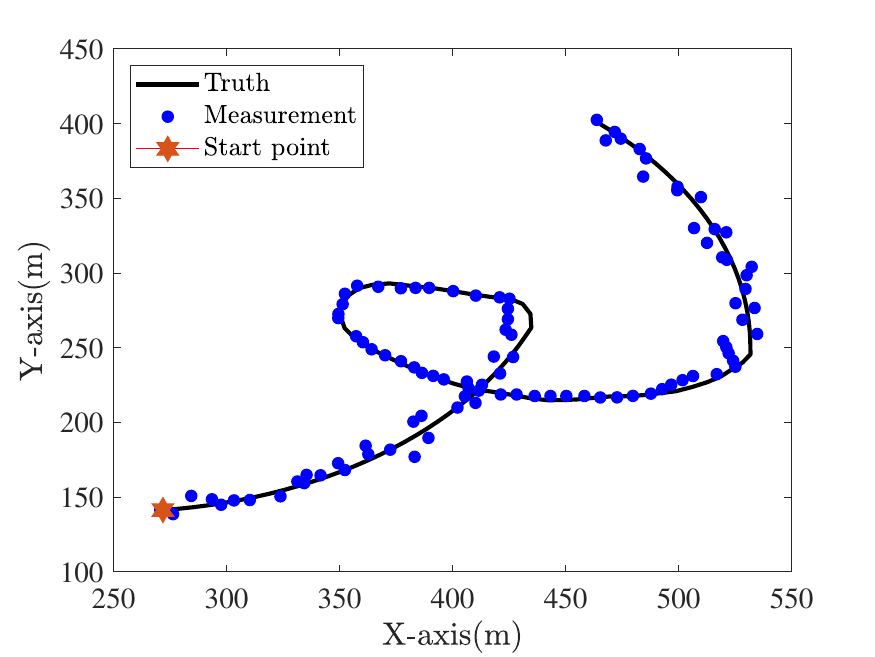}} 
    \caption{Trajectories of targets and radar measurements in real-world scenarios.}
    \label{S4::RMSE}
\end{figure}

Three different maneuvering target tracking applications with real-world 2D surveillance radar datasets are used to evaluate the performance of different VB-based adaptive filtering methods. %, including aerial target tracking, vehicle target tracking, and ship target tracking.
\begin{itemize}
    \item \textbf{R1}\cite{10247583}: The aerial target trajectory is shown in Fig.~\ref{fig::R1}, which consists of six different segments or maneuver modes. These segments include segment-A~(CV), segment-B~(left CT), segment C~(CV), segment D~(right CT), segment E~(CA), and segment-F~(figure-eight flight pattern). 
    \item \textbf{R2}\cite{10247583}: The vehicle target trajectory is shown in Fig.~\ref{fig::R2} which consists of round-trip maneuver.
    \item \textbf{R3}\cite{10247583}: The ship target trajectory is shown in Fig.~\ref{fig::R3} which consists of coordinated turn maneuver.
\end{itemize}

\begin{figure}[htbp]
    \centering
    \subfloat[\textbf{R1}]{\label{fig::RMSE-R1} \includegraphics[scale=0.45]{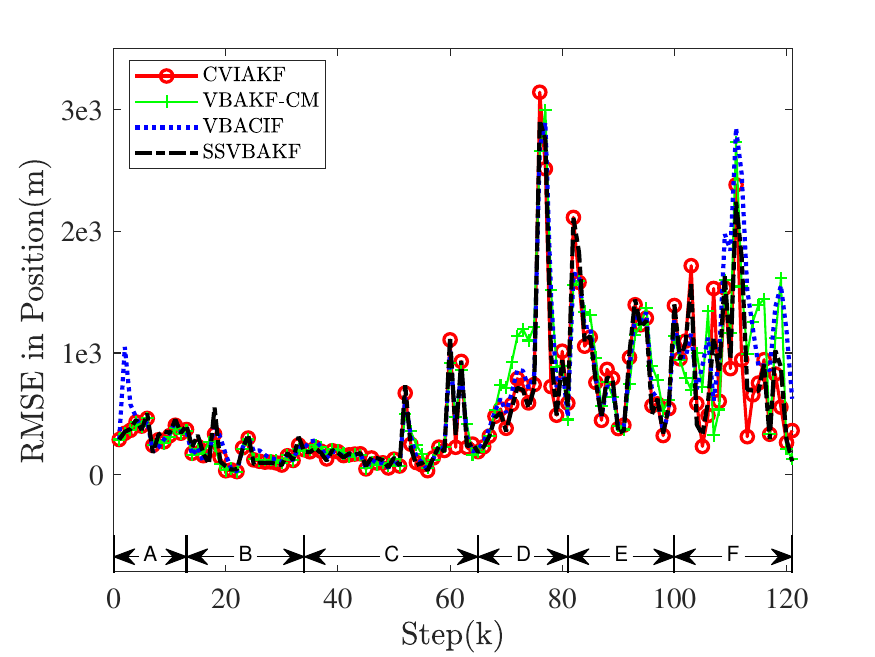}} \\ 
    \subfloat[\textbf{R2}]{\label{fig::RMSE-R2} \includegraphics[scale=0.45]{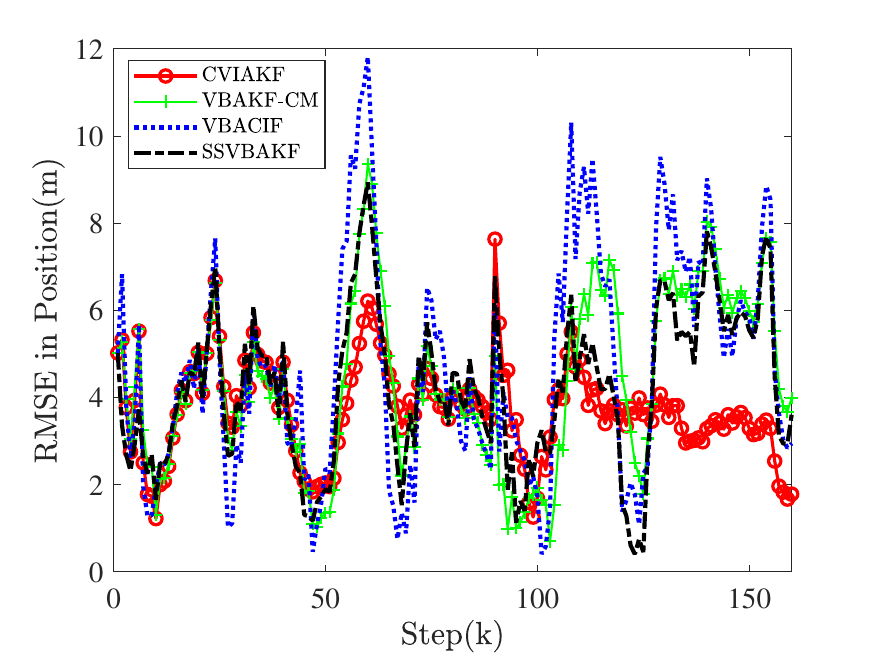}} \\
    \subfloat[\textbf{R3}]{\label{fig::RMSE-R3} \includegraphics[scale=0.45]{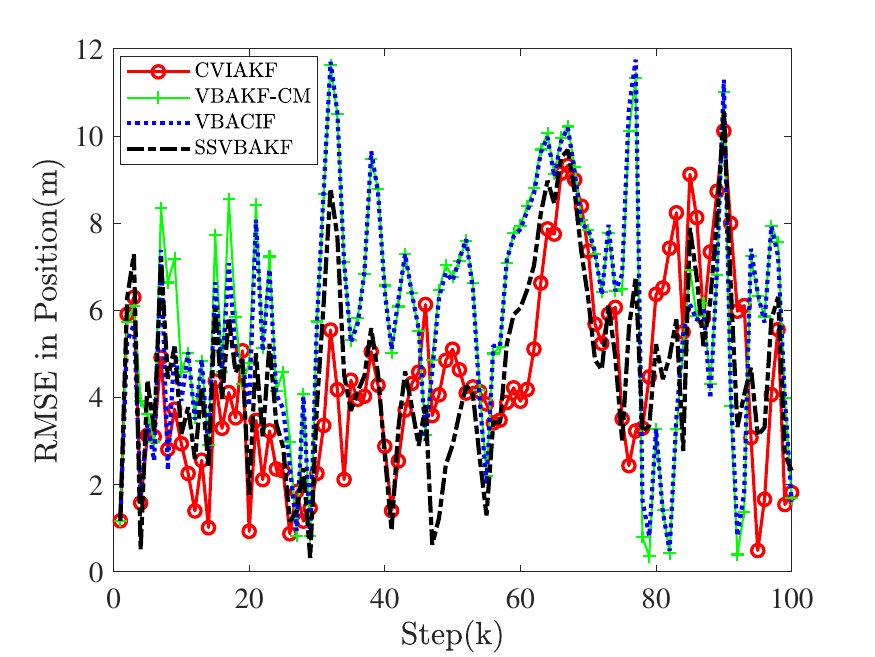}}
    \caption{Position RSME comparison for real-world scenarios}
    \label{R::RMSE}
\end{figure}

The radar sampling period $T$ for \textbf{R1}, \textbf{R2}, and \textbf{R3} are 10s, 2s, and 3s, respectively. The algorithms are configured with the same parameters as the synthetic data scenarios. The results are shown as follows. Fig.\ref{R::RMSE} and Table.\ref{table::ARMSE_P for real data} depict the RMSE and ARMSE for different real-world datasets, respectively. It is seen that CVIAKF has the best tracking performance for all real scenarios, followed by SSVBAKF and VBAKF-CM, and VBACIF is the worst. In particular, Fig.~\ref{fig::RMSE-R1} and Table.~\ref{table::ARMSE_P of segment for R1} show the segment analysis of RMSE for scenario \textbf{R1}. When the target moves with CV modes or with small turning maneuvers, all adaptive Kalman filtering methods perform well. However, VBAEKF-CM and CVIAKF have proven to be the most effective. When the target executes a large turning maneuver (segments D and F) or experiences acceleration (segment E), the filtering performance of all methods degrades. Nonetheless, CVIAKF exhibits relatively robust performance for different maneuvers.
In contrast to scenario \textbf{R1}, the targets in scenarios \textbf{R2} and \textbf{R3} exhibit a more consistent movement pattern characterized by straight lines or turns.
In these scenarios, all adaptive Kalman filtering methods can produce satisfactory estimation results.
The average iterations for VBACIF, VBAKF-CM, SSVBAKF, and CVIAKF are 5, 5, 914, 183 for scenario \textbf{R1}, 4, 3, 6, 14 for scenario \textbf{R2}, and 5, 6, 590, 180 for scenario \textbf{R3}, respectively. 

\begin{table}[htbp]
	\centering  
	\caption{ARMSE comparison of positions for real-world scenarios~(units: m)} 
	\label{table::ARMSE_P for real data} 
	\begin{tabular}{c|cccc}  
		\hline \hline % 表格的横线
		\textbf{Scenario} & \textbf{VBACIF}& \textbf{VBAKF-CM}& \textbf{SSVBAKF} & \textbf{CVIAKF}\\ \hline	
        \textbf{R1} & 628.10 & 597.52 &548.25 & \textbf{\textcolor{blue}{531.54}}
        \\ \hline
         \textbf{R2} & 4.72 & 4.38 & 4.22& \textbf{\textcolor{blue}{3.71}}
        \\ \hline
         \textbf{R3} & 5.80 & 5.96 & 4.64 & \textbf{\textcolor{blue}{4.42}}
        \\ \hline
	\end{tabular}
\end{table}

\begin{table}[htbp]
	\centering  
	\caption{ARMSE comparison of positions of different segments for scenario \textbf{R1}~(units: m)} 
	\label{table::ARMSE_P of segment for R1} 
	\begin{tabular}{c|cccc}  
		\hline \hline % 表格的横线
		\textbf{Segment} & \textbf{VBACIF}& \textbf{VBAKF-CM}& \textbf{SSVBAKF}& \textbf{CVIAKF}\\ \hline	
  
        A : [0, 13)  &419.99 & \textbf{\textcolor{blue}{341.10}}  &361.40 & 346.06
        \\ 
         B : [13, 34) &186.07 & \textbf{\textcolor{blue}{156.18}} &177.24 & 159.45
        \\ 
         C : [34, 65) &237.36 & 238.56 &234.03 & \textbf{\textcolor{blue}{226.87}}
        \\ 
         D : [65, 81) &943.45 & 1077.90  & \textbf{\textcolor{blue}{835.06}} & 840.71
        \\ 
         E : [81, 100) &\textbf{\textcolor{blue}{877.71}} & 908.06  &904.38 & 893.37
        \\ 
         F : [100, 121] &1269.20 & 1046.80 & 930.95&\textbf{\textcolor{blue}{878.05}}
        \\ \hline
	\end{tabular}
\end{table}

\section{Conclusion}
We have proposed CVIAKF for both linear and nonlinear state space models. 
Unlike the existing adaptive Kalman filters that directly apply variational inference in natural-parameter space, CVIAKF  
conducts optimization in the expectation-parameter space, leading to a faster and simpler solution, especially for nonlinear state estimation. Meanwhile, CVIAKF divides the ELBO optimization into conjugate parts and non-conjugate parts for nonlinear dynamic models, for which the conjugate computations and stochastic mirror-descent are applied, respectively. The reparameterization trick is used to reduce the variance of stochastic gradients of the non-conjugate parts, which can further improve the estimation accuracy. The positive-definite constraint in the stochastic mirror-descent method has been considered for enhancing numerical stability. The CVIAKF method has been validated with synthetic and real-world datasets for maneuvering target tracking.

\section*{APPENDIX A} 
For linear models, the updates of natural parameters $\bm{\lambda}_k^{\bm{x}}$, $\bm{\lambda}_k^{\bm{P}}$ and $\bm{\lambda}_k^{\bm{R}}$ can be obtained directly by setting the deviations of ELBO $\mathcal{\bar B}(\bm{\mu}_k)$ equal to zeros. For simplicity, here and thereafter, we denote the expectation $\mathbb{E}_{q(\bm{x})}[\bm{x}]$ as $\mathbb{E}[\bm{x}]$.

\begin{figure*} [!tp]
\normalsize
\setcounter{MYtempeqncnt1}{\value{equation}}
\setcounter{equation}{32}
\begin{equation}
    \label{B_x}
    \begin{split}
        \mathcal{\bar B}(\bm{\mu}_k^{\bm{x}}) =& \mathbb{E}\left[-0.5(\bm{y}_k - \bm{H}_k\bm {x}_k)^{\top}\bm{R}_k^{-1} (\bm{y}_k - \bm{H}_k\bm{x}_k) \right]
        +\mathbb{E} \left[ -0.5(\bm{x}_k - \hat{\bm{x}}_{k|k-1})^{\top}\bm{P}_{k|k-1}^{-1} (\bm{x}_k - \hat{\bm{x}}_{k|k-1}) \right]
        \\
          &-\mathbb{E} \left[ -0.5(\bm{x}_k - \hat{\bm{x}}_{k|k})^{\top}\bm{P}_{k|k}^{-1} (\bm{x}_k - \hat{\bm{x}}_{k|k}) \right] \\
          =&\mathrm{Tr}\left\{\big(\bm{H}_k^{\top}\mathbb{E}[\bm{R}_k^{-1}]\bm{y}_k + \mathbb{E}[\bm{P}_{k|k-1}^{-1}]\hat{\bm{x}}_{k|k-1} - \bm{P}_{k|k}^{-1} \hat{\bm{x}}_{k|k} \big)(\bm{\mu}_k^{\bm{x}}(1))^{\top}\right\}
     \\
     &-0.5\mathrm{Tr}\left\{\big(\bm{H}_k^{\top}\mathbb{E}[\bm{R}_k^{-1}]\bm{H}_k + \mathbb{E}[\bm{P}_{k|k-1}^{-1}] - \bm{P}_{k|k}^{-1}\big) \bm{\mu}_k^{\bm{x}}(2) \right\}
        \end{split}
    \end{equation} 
\hrulefill
\setcounter{equation}{\value{MYtempeqncnt1}}
\end{figure*}
\addtocounter{equation}{1}

\emph{1. Derivations of $\bm{\lambda}_k^{\bm{x}}$}: Omitting the terms that are independent of $\bm{x}_k$, the simplified ELBO $\mathcal{\bar B}(\bm{\mu}_k)$, denoted by $\mathcal{\bar B}(\bm{\mu}_k^{\bm{x}})$, can be written as \eqref{B_x}, where $\bm{\mu}_k^{\bm{x}}(1) =\mathbb{E}[\bm{x}_k] $ and $\bm{\mu}_k^{\bm{x}}(2) =  \mathbb{E}[\bm{x}_k \bm{x}_k^{\top}]$.

Let the gradients of \eqref{B_x} w.r.t. $\bm{\mu}_k^{\bm{x}}$ be equal to zeros, the  variational parameters $\bm{\hat x}_k$ and $\bm{P}_{k|k}$ can be updated as
\begin{equation}
     \begin{split}
        \bm{P}_{k|k}^{-1} \hat{\bm{x}}_{k|k} = \,& \mathbb{E}[\bm{P}_{k|k-1}^{-1}]\hat{\bm{x}}_{k|k-1} + \bm{H}_k^{\top}\mathbb{E}[\bm{R}_k^{-1}]\bm{y}_k,
        \\
        \bm{P}_{k|k}^{-1} =  \,& \mathbb{E}[\bm{P}_{k|k-1}^{-1}]  + \bm{H}_k^{\top}\mathbb{E}[\bm{R}_k^{-1}]\bm{H}_k.
    \end{split}
\end{equation}

\begin{figure*} [!tp]
\normalsize
\setcounter{MYtempeqncnt2}{\value{equation}}
\setcounter{equation}{34}
\begin{equation}
\label{B_P}
    \begin{split}
        \mathcal{\bar B}(\bm{\mu}_k^{\bm{P}}) = & \mathbb{E} \left[ -0.5(\bm{x}_k - \hat{\bm{x}}_{k|k-1})^{\top}\bm{P}_{k|k-1}^{-1} (\bm{x}_k - \hat{\bm{x}}_{k|k-1})- 0.5\log\left| \bm{P}_{k|k-1}\right| \right] 
        \\
        &+  \mathbb{E} \left[-0.5(\hat{u}_{k|k-1} + n + 1)\log\left| \bm{P}_{k|k-1}\right| -0.5\mathrm{Tr}\{\bm{U}_{k|k-1}\bm{P}_{k|k-1}^{-1}\} \right]
        \\
        &+ \mathbb{E} \left[0.5(\hat{u}_{k|k} + n + 1)\log\left|\bm{P}_{k|k-1}\right|+0.5\mathrm{Tr}\{\bm{U}_{k|k}\bm{P}_{k|k-1}^{-1}\} \right] \\
        &= 0.5(\hat{u}_{k|k} - \hat{u}_{k|k-1} - 1)\bm{\mu}_k^{\bm{P}}(1) + 0.5\mathrm{Tr}\left\{\left(\bm{U}_{k|k} -\bm{C}_k - \bm{U}_{k|k-1} \right) \bm{\mu}_k^{\bm{P}}(2) \right\}
    \end{split}
\end{equation}
\hrulefill
\setcounter{equation}{\value{MYtempeqncnt2}}
\end{figure*}
\addtocounter{equation}{1}

\emph{2. Derivations of $\bm{\lambda}_k^{\bm{P}}$}: 
Omitting the terms that are independent of $\bm{P}_{k|k-1}$, the simplified ELBO $\mathcal{\bar B}(\bm{\mu}_k)$, denoted by $\mathcal{\bar B}(\bm{\mu}_k^{\bm{P}})$, can be written as \eqref{B_P}, where
$\bm{\mu}_k^{\bm{P}}(1) = \mathbb{E}[\log| \bm{P}_{k|k-1}|] $, $\bm{\mu}_k^{\bm{P}}(2) =  \mathbb{E}[\bm{P}_{k|k-1}^{-1}]$, and $\bm{C}_k \triangleq \mathbb{E}[(\bm{x}_k - \hat{\bm{x}}_{k|k-1})(\bm{x}_k - \hat{\bm{x}}_{k|k-1})^{\top}] = (\hat{\bm{x}}_{k|k} - \hat{\bm{x}}_{k|k-1})(\hat{\bm{x}}_{k|k} - \hat{\bm{x}}_{k|k-1})^{\top} + \bm{P}_{k|k}$.

Let the gradients of \eqref{B_P} w.r.t. $\bm{\mu}_k^{\bm{P}}$ be equal to zeros, the variational parameters $u_{k|k}$ and $\bm{U}_{k|k}$ can be updated as
\begin{equation}
\label{eq::update hyper-parameter of P}
    \begin{split}
        \hat{u}_{k|k} =& \hat{u}_{k|k-1} + 1,        \\
        \bm{U}_{k|k} = & \bm{U}_{k|k-1} + \bm{C}_k.
    \end{split}
\end{equation}

\emph{3. Derivations of $\bm{\lambda}_k^{\bm{R}}$}: Omitting the terms that are independent of $\bm{R}_{k}$, the simplified ELBO $\mathcal{\bar B}(\bm{\mu}_k)$, denoted by $\mathcal{\bar B}(\bm{\mu}_k^{\bm{R}})$, can be written as \eqref{B_R}, where $\bm{\mu}_k^{\bm{R}}(1) = \mathbb{E}[\log| \bm{R}_{k}|] $, $\bm{\mu}_k^{\bm{R}}(2) =  \mathbb{E}[\bm{R}_{k}^{-1}]$, and $\bm{A}_k \triangleq \mathbb{E}[(\bm{y}_k - \bm{H}_k\bm{x}_k)(\bm{y}_k - \bm{H}_k\bm{x}_k)^{\top}] = (\bm{y}_k - \bm{H}_k\hat{\bm{x}}_{k|k})(\bm{y}_k - \bm{H}_k\hat{\bm{x}}_{k|k})^{\top} + \bm{H}_k\bm{P}_{k|k}\bm{H}_k^{\top}$.

\begin{figure*} [!tp]
\normalsize
\setcounter{MYtempeqncnt3}{\value{equation}}
\setcounter{equation}{36}
\begin{equation}
\label{B_R}
    \begin{split}
        \mathcal{\bar B}(\bm{\mu}_k^{\bm{R}}) = & \mathbb{E} \left[ -0.5(\bm{y}_k - \bm{H}_k\bm{x}_k )^{\top}\bm{R}_k^{-1} (\bm{y}_k - \bm{H}_k\bm{x}_k) - 0.5\log\left| \bm{R}_k\right| \right] 
        \\
        &- \mathbb{E} \left[0.5(\hat{v}_{k|k-1} + m + 1)\log\left| \bm{R}_k\right| + 0.5\mathrm{Tr}\{\bm{V}_{k|k-1}\bm{R}_k^{-1}\}  \right]
        \\
        &+ \mathbb{E} \left[0.5(\hat{v}_{k|k} + m + 1)\log\left| \bm{R}_k\right| + 0.5 \mathrm{Tr}\{\bm{V}_{k|k}\bm{R}_k^{-1}\}\right] \\
         =& 0.5(\hat{v}_{k|k} - \hat{v}_{k|k-1} - 1)\bm{\mu}_k^{\bm{R}}(1) + 0.5\mathrm{Tr}\left\{\left(\bm{V}_{k|k} -\bm{A}_k - \bm{V}_{k|k-1} \right) \bm{\mu}_k^{\bm{R}}(2) \right\}
     \end{split}
\end{equation}
\hrulefill
\setcounter{equation}{\value{MYtempeqncnt3}}
\end{figure*}
\addtocounter{equation}{1}

Let the gradients of \eqref{B_R} w.r.t. $\bm{\mu}_k^{\bm{R}}$ be equal to zeros, the variational parameters $v_{k|k}$ and $\bm{V}_{k|k}$ can be updated as
\begin{equation}
\label{eq::update hyper-parameter of R}
    \begin{split}
        \hat{v}_{k|k} =& \hat{v}_{k|k-1} + 1,        \\
        \bm{V}_{k|k} = & \bm{V}_{k|k-1} + \bm{A}_k.
    \end{split}
\end{equation}

The proof of \textbf{Theorem 1} has been completed. $\hfill \Large\square$

 \section*{APPENDIX B} 
For linear models, the update of the natural parameters $\bm{\lambda}_k^x$ \eqref{x in linear case} can be simplified as
 \begin{align}
     \label{eq::CVIAKF-invP}
         \bm{P}_{k|k}^{-1} =\,& \mathbb{E}[\bm{P}_{k|k-1}^{-1}] + \bm{H}_k^{\top} \mathbb{E}[\bm{R}_k^{-1}]\bm{H}_k,
         \\
         \label{eq::CVIAKF-x}
         \hat{\bm{x}}_{k|k} = \,& \bm{P}_{k|k}\left[\mathbb{E}[\bm{P}_{k|k-1}^{-1}]\hat{\bm{x}}_{k|k-1} + \bm{H}_{k}^{\top} \mathbb{E}[\bm{R}_k^{-1}]\bm{y}_k\right].
 \end{align}
Inverting both sides of \eqref{eq::CVIAKF-invP} and employing the Sherman-Morrison-Woodbury Lemma, we have 
 \begin{equation}
 \label{eq::hyperparameter P}
     \begin{split}
         \bm{P}_{k|k} = \,& \left[ \mathbb{E}[\bm{P}_{k|k-1}^{-1}] + \bm{H}_k^{\top} \mathbb{E}[\bm{R}_k^{-1}]\bm{H}_k\right]^{-1}
         \\
         =\,& \tilde{\bm{P}}_{k|k-1} - \bm{K}_k\bm{H}_k \tilde{\bm{P}}_{k|k-1}
         \\
         =\,& (\bm{\mathrm{I}}_n - \bm{K}_k\bm{H}_k)\tilde{\bm{P}}_{k|k-1}
     \end{split}
 \end{equation}
 with $\tilde{\bm{P}}_{k|k-1}^{-1} \triangleq \mathbb{E}[\bm{P}_{k|k-1}^{-1}] $, $\Tilde{\bm{R}}_k^{-1} \triangleq \mathbb{E}[\bm{R}_k^{-1}]$, and the filter gain $\bm{K}_k = \tilde{\bm{P}}_{k|k-1} \bm{H}_k^{\top}(\tilde{\bm{R}}_k + \bm{H}_k \tilde{\bm{P}}_{k|k-1} \bm{H}_k^{\top})^{-1}$.

By the fact that the following equation holds
 \begin{equation}
     \begin{split}
         \bm{K}\Tilde{\bm{R}}_k = \,& \bm{K}(\Tilde{\bm{R}}_k + \bm{H}_k \Tilde{\bm{P}}_{k|k-1}\bm{H}^{\top})- \bm{K}\bm{H}_k \Tilde{\bm{P}}_{k|k-1}\bm{H}_k^{\top}
         \\
         =\,& \tilde{\bm{P}}_{k|k-1} \bm{H}_k^{\top} - \bm{K}\bm{H}_k \Tilde{\bm{P}}_{k|k-1}\bm{H}_k^{\top}
         \\
         =\,& (\bm{\mathrm{I}}_n - \bm{K}\bm{H}_k)\Tilde{\bm{P}}_{k|k-1}\bm{H}_k^{\top}
         \\
         =\,& \bm{P}_{k|k}\bm{H}_k^{\top},
     \end{split}
 \end{equation}
the filter gain can be rewritten as $\bm{K} = \bm{P}_{k|k} \bm{H}_k^{\top} \Tilde{\bm{R}}_k^{-1}$.

 Therefore, \eqref{eq::CVIAKF-x} can be expressed as
 \begin{equation}
 \label{eq::hyperparameter x}
     \begin{split}
         \hat{\bm{x}}_{k|k} =\,& \bm{P}_{k|k} \Tilde{\bm{P}}_{k|k-1}^{-1}\hat{\bm{x}}_{k|k-1} + \bm{P}_{k|k}\bm{H}_k^{\top} \Tilde{\bm{R}}_{k}^{-1}\bm{y}_k
         \\
         =\,& \hat{\bm{x}}_{k|k-1}-\bm{K}\bm{H}_k \hat{\bm{x}}_{k|k-1} + \bm{K}\bm{y}_k
         \\
         =\,& \hat{\bm{x}}_{k|k-1} +\bm{K}(\bm{y}_k - \bm{H}_k \hat{\bm{x}}_{k|k-1}).
     \end{split}
 \end{equation}

From the proof of \textbf{Theorem 1},  the updated natural parameters of $\bm{P}_{k|k-1}$ and $\bm{R}_k$ can be simplified as \eqref{eq::update hyper-parameter of P} and \eqref{eq::update hyper-parameter of R}, respectively. Combining \eqref{eq::hyperparameter P} with \eqref{eq::hyperparameter x}, it is demonstrated that CVIAKF is equivalent to VBAKF.

When $\bm{Q}_k$ and $\bm{R}_k$ are known, the predicted state $\bm{P}_{k|k-1}$ and $\bm{R}_k$ can be obtained accurately.
Then, $\mathbb{E}[\bm{P}_{k|k-1}^{-1}] = \bm{P}_{k|k-1}^{-1}$ and $\mathbb{E}[\bm{R}_{k}^{-1}] = \bm{R}_{k}^{-1}$.
The updated natural parameters $\bm{\lambda}_k^{x}$ \eqref{x in linear case} can be simplified as
\begin{equation}
\label{eq::information filter}
    \begin{split}
        \bm{P}_{k|k}^{-1} =\,& \bm{P}_{k|k-1}^{-1} + \bm{H}_k^{\top} \bm{R}_k^{-1}\bm{H}_k,
         \\
         \bm{P}_{k|k}^{-1}\hat{\bm{x}}_{k|k} = \,&  \bm{P}_{k|k-1}^{-1}\hat{\bm{x}}_{k|k-1} + \bm{H}_{k}^{\top} \bm{R}_k^{-1}\bm{y}_k.
    \end{split}
\end{equation}
The update of simplified parameters \eqref{eq::information filter} is a standard information filter, where $\bm{P}_{k|k}^{-1}$ and $\bm{P}_{k|k}^{-1}\hat{\bm{x}}_{k|k}$ are respectively defined as the information matrix and information vector.
The predicted information matrix and predicted information vector are $\bm{P}_{k|k-1}^{-1}$ and $\bm{P}_{k|k-1}^{-1}\hat{\bm{x}}_{k|k-1}$.
The measurement covariance is $\bm{H}_k^{\top} \bm{R}_k^{-1}\bm{H}_k$, and the measurement vector is $\bm{H}_{k}^{\top} \bm{R}_k^{-1}\bm{y}_k$.

The proof of \textbf{Proposition 1} has been completed. $\hfill \Large\square$

\section*{APPENDIX C} 
For nonlinear models, it is intractable to optimize the ELBO objective directly. One should resort to the stochastic mirror descent method. 

\begin{figure*} [!tp]
\normalsize
\setcounter{MYtempeqncnt4}{\value{equation}}
\setcounter{equation}{44}
\begin{equation}
    \label{B_x-n}
    \begin{split}
        \mathcal{\bar B}(\bm{\mu}_k^{\bm{x}}) =& -0.5\underbrace{\mathbb{E}\left[(\bm{y}_k - \bm{h}_k(\bm {x}_k))^{\top}\bm{R}_k^{-1} (\bm{y}_k - \bm{h}_k(\bm{x}_k)) \right]}_{\mathcal{D}(\bm{\mu}_k^{\bm{x}})}      \\
          &-0.5\underbrace{\mathbb{E} \left[(\bm{x}_k - \hat{\bm{x}}_{k|k-1})^{\top}\bm{P}_{k|k-1}^{-1} (\bm{x}_k - \hat{\bm{x}}_{k|k-1}) - (\bm{x}_k - \hat{\bm{x}}_{k|k})^{\top}\bm{P}_{k|k}^{-1} (\bm{x}_k - \hat{\bm{x}}_{k|k})\right]}_{\mathcal{E}(\bm{\mu}_k^{\bm{x}})}
        \end{split}
\end{equation} 
\hrulefill
\setcounter{equation}{\value{MYtempeqncnt4}}
\end{figure*}
\addtocounter{equation}{1}

\emph{1. Derivations of $\bm{\lambda}_k^{\bm{x}}$}: Omitting the terms that are independent of $\bm{x}_k$, the simplified ELBO $\mathcal{\bar B}(\bm{\mu}_k)$, denoted by $\mathcal{\bar B}(\bm{\mu}_k^{\bm{x}})$, can be written as \eqref{B_x-n}, 
where $\mathcal{E}(\bm{\mu}_k^{\bm{x}})$ and $\mathcal{D}(\bm{\mu}_k^{\bm{x}})$ are the conjugate part and non-conjugate part of $\mathcal{\bar B}(\bm{\mu}_k^{\bm{x}})$, respectively.

The gradient of the ELBO $\mathcal{\bar B}(\bm{\mu}_k^{\bm{x}})$ w.r.t. expectation parameters $\bm{\mu}_k^{\bm{x}}$ can be expressed as 
\begin{equation}
\label{x in nonlinear case}
    \nabla_{\bm{\mu}_k^{\bm{x}}} \mathcal{\bar{B}}(\bm{\mu}_k^{\bm{x}}) = -0.5\nabla_{\bm{\mu}_k^{\bm{x}}} \mathcal{E}(\bm{\mu}_k^{\bm{x}}) -0.5 \nabla_{\bm{\mu}_k^{\bm{x}}} \mathcal{D}(\bm{\mu}_k^{\bm{x}}). 
\end{equation}

\begin{figure*} [!tp]
\normalsize
\setcounter{MYtempeqncnt5}{\value{equation}}
\setcounter{equation}{46}
\begin{equation}
\label{conjugate_x}
    \begin{split}
        \nabla_{\bm{\mu}_k^{\bm{x}}} \mathcal{E}(\bm{\mu}_k^{\bm{x}})
        = \nabla_{\bm{\mu}_k^{\bm{x}}} \mathrm{Tr}\left\{2\big(-\mathbb{E}[\bm{P}_{k|k-1}^{-1}]\hat{\bm{x}}_{k|k-1} + \bm{P}_{k|k}^{-1} \hat{\bm{x}}_{k|k}\big) (\bm{\mu}_k^{\bm{x}}(1))^{\top}\right\} 
        + \nabla_{\bm{\mu}_k^{\bm{x}}}\mathrm{Tr}\left\{\big(\mathbb{E}[\bm{P}_{k|k-1}^{-1}] -\bm{P}_{k|k}^{-1}\big)\bm{\mu}_k^{\bm{x}}(2)\right\}
    \end{split}
\end{equation}
\hrulefill
\setcounter{equation}{\value{MYtempeqncnt5}}
\end{figure*}
\addtocounter{equation}{1}

For the conjugate part $\mathcal{E}(\bm{\mu}_k^{\bm{x}})$, its gradient can be computed as \eqref{conjugate_x}, where $\bm{\mu}_k^{\bm{x}} = [\bm{\mu}_k^{\bm{x}}(1), \bm{\mu}_k^{\bm{x}}(2)]^{\top}$ with $\bm{\mu}_k^{\bm{x}}(1) =\mathbb{E}[\bm{x}_k] $ and $\bm{\mu}_k^{\bm{x}}(2) =  \mathbb{E}[\bm{x}_k \bm{x}_k^{\top}]$.
Then, the gradient of conjugate part $\nabla_{\bm{\mu}_k^{\bm{x}}} \mathcal{E}(\bm{\mu}_k^{\bm{x}})$ can be calculated in analytical form as 
\begin{equation}
\label{conjugate_gradient}
    \begin{split}
        \nabla_{\bm{\mu}_k^{\bm{x}}(1)} \mathcal{E}(\bm{\mu}_k^{\bm{x}}) = \,& -2\mathbb{E}[\bm{P}_{k|k-1}^{-1}] \hat{\bm{x}}_{k|k-1} + 2\bm{P}_{k|k}^{-1} \hat{\bm{x}}_{k|k},
       \\
        \nabla_{\bm{\mu}_k^{\bm{x}}(2)} \mathcal{E}(\bm{\mu}_k^{\bm{x}})= \,& \mathbb{E}[\bm{P}_{k|k-1}^{-1}] - \bm{P}_{k|k}^{-1}.
    \end{split}
\end{equation}

\begin{figure*} [!tp]
\normalsize
\setcounter{MYtempeqncnt6}{\value{equation}}
\setcounter{equation}{48}
\begin{equation}
\label{non-conjugate_x}
    \begin{split}
        \nabla_{\bm{\mu}_k^{\bm{x}}} \mathcal{D}(\bm{\mu}_k^{\bm{x}})
        = \nabla_{\bm{\mu}_k^{\bm{x}}} \mathrm{Tr}\left\{\mathbb{E}\left[\bm{R}_k^{-1}\right]\mathbb{E}\left[(\bm{y}_k - \bm{h}_k(\bm {x}_k))(\bm{y}_k - \bm{h}_k(\bm {x}_k))^{\top}\right]\right\}
    \end{split}
\end{equation}
\hrulefill
\setcounter{equation}{\value{MYtempeqncnt6}}
\end{figure*}
\addtocounter{equation}{1}

For the non-conjugate part $\mathcal{D}(\bm{\mu}_k^{\bm{x}})$, its gradient can be computed as \eqref{non-conjugate_x}.
It is intractable to calculate the above gradient due to the nonlinear measurement function $\bm{h}_k(\bm{x}_k)$. By using the chain rule, the gradient w.r.t. multivariate Gaussian distribution $\bm{x}_k \sim \bm{\mathrm{N}}(\bm{x}_k|\bm{\hat x}_{k|k}, \bm{P}_{k|k})$ can be rewritten as 
\begin{equation}
\label{gradient_transform}
\begin{split}
        \nabla_{\bm{\mu}_k^{\bm{x}}(1)} \mathcal{D}(\bm{\mu}_k^{\bm{x}}) =
        \,&\nabla_{\hat{\bm{x}}_{k|k}} \mathcal{D}(\bm{\mu}_k^{\bm{x}}) - 2 \nabla_{\bm{P}_{k|k}} \mathcal{D}(\bm{\mu}_k^{\bm{x}})\hat{\bm{x}}_{k|k}, 
        \\
        \nabla_{\bm{\mu}_k^{\bm{x}}(2)} \mathcal{D}(\bm{\mu}_k^{\bm{x}}) = & \nabla_{\bm{P}_{k|k}} \mathcal{D}(\bm{\mu}_k^{\bm{x}}).
\end{split}
\end{equation}

Let $\varphi(\bm{x}_k) =(\bm{y}_k - \bm{h}_k(\bm{x}_k))^{\top} \bm{R}_k^{-1}(\bm{y}_k - \bm{h}_k(\bm{x}_k))$.
Using the Monte Carlo integration, and Bonnet's theorem and Price's theorem~\cite{murphy2023probabilistic}, the gradients in \eqref{gradient_transform} can be approximated by \eqref{D_x_non-conjugate}-\eqref{D_P_non-conjugate}, where $\bm{x}_k^{s} \overset{\mathrm{iid}}{\sim} \bm{\mathrm{N}}(\bm{x}_k|\bm{\hat x}_{k|k}, \bm{P}_{k|k})$, $s = 1,2,\cdots,S$ are the sampling particles with $S$ being the number of samples. $\bm{H}_k(\bm{x}_k) \in \mathbb{R}^{m \times n}$ denotes the Jacobian matrix of $\bm{h}_k(\bm{x}_k)$. Let $\mathcal{H}_i(\bm{x}_k) \in \mathbb{R}^{1\times n}$ denote the $i$th row of $\bm{H}_k(\bm{x}_k)$, then
$\bm{G}_i(\bm{x}_k) \in \mathbb{R}^{n\times n}$ is the Jacobian matrix of $\mathcal{H}_i(\bm{x}_k)$. $B_i(\bm{x}_k)$ denotes the $i$th element of the vector $\bm{R}_k^{-1}(\bm{y}_k - \bm{h}_k(\bm{x}_k))$.

\begin{figure*} [!tp]
\normalsize
\setcounter{MYtempeqncnt7}{\value{equation}}
\setcounter{equation}{50}
\begin{equation}
\label{D_x_non-conjugate}
    \begin{split}
       \nabla_{\hat{\bm{x}}_{k|k}} \mathcal{D}(\bm{\mu}_k^{\bm{x}})= \mathbb{E}\left[\nabla_{\bm{x}_{k}} \varphi(\bm{x}_k)\right] = -2\mathbb{E} \left[ \bm{H}^{\top}_k(\bm{x}_k)\bm{R}_k^{-1} (\bm{y}_k - \bm{h}_k(\bm {x}_k)) \right] 
        \approx -\frac{2}{S}\sum_{s = 1} ^{S} \bm{H}^{\top}_k(\bm{x}_k^s)\mathbb{E}[\bm{R}_k^{-1}](\bm{y}_k - \bm{h}_k(\bm {x}_k^s))
    \end{split}
\end{equation}
\hrulefill
\begin{equation}
\label{D_P_non-conjugate}
    \begin{split}
       \nabla_{\bm{P}_{k|k}} \mathcal{D}(\bm{\mu}_k^{\bm{x}}) =& 0.5\mathbb{E}\left[\nabla^2_{\bm{x}_{k}} \varphi(\bm{x}_k)\right] 
       = \mathbb{E}\left[\bm{H}^{\top}_k(\bm{x}_k)\bm{R}_k^{-1}\bm{H}_k(\bm{x}_k)   - \sum_{i=1}^m\bm{G}_i^{\top} (\bm{x}_k)B_i(\bm{x}_k)\right] 
       \\
        \approx& \frac{1}{S}\sum_{s = 1} ^{S} \left[\bm{H}^{\top}_k(\bm{x}_k^s)\mathbb{E}[\bm{R}_k^{-1}]\bm{H}_k(\bm{x}_k^s) - \sum_{i=1}^m\bm{G}_i^{\top} (\bm{x}^s_k)B_i(\bm{x}^s_k) \right]
    \end{split}
\end{equation}
\hrulefill
\setcounter{equation}{\value{MYtempeqncnt7}}
\end{figure*}
\addtocounter{equation}{2}

In practice, the gradients in \eqref{D_x_non-conjugate}-\eqref{D_P_non-conjugate} approximated by Monte Carlo integration may have high variance. To address this issue, we employ the reparameterization trick \cite{mohamed2020monte} to reduce the estimation variance. 

Assume that the random variable $\bm{x}_k$ can be expressed as the noise term $\bm{\epsilon}$ and deterministic terms $\bm{\theta}_k$, the reparameterization trick involves sampling from a noise distribution $q_0(\bm{\epsilon}_k)$, which is independent of $\bm{\theta}_k$, and subsequently transforming to $\bm{x}_k$ through a deterministic and differentiable function $\bm{x}_k = \bm{g}(\bm{\epsilon}_k)$. 
For Gaussian random variable $\bm{x}_k$, instead of sampling $\bm{x}_k \sim \bm{\mathrm{N}}(\bm{x}_k|\bm{\hat x}_{k|k}, \bm{P}_{k|k})$, we can sample $\bm{\epsilon}_k \sim \bm{\mathrm{N}}(\bm{\mathrm{0}}, \bm{\mathrm{I}}_{n})$ and compute $\bm{x}_k = \bm{g}(\bm{\epsilon}_k) = \hat{\bm{x}}_{k|k} + \bm{L}_k\bm{\epsilon}_k$ with $\bm{L}_k\bm{L}_k^{\top} = \bm{P}_{k|k}$. Then, the reparameterization gradients are computed by
\begin{equation}
\label{RP_non-conjugate_x}
    \begin{split}
      \nabla_{\hat{\bm{x}}_{k|k}} \mathcal{D}(\bm{\mu}_k^{\bm{x}})= &\mathbb{E}\left[\nabla_{\bm{g}(\bm{\epsilon}_k)}\varphi(\bm{x}_k) \nabla_{\hat{\bm{x}}_{k|k}}\bm{g}(\bm{\epsilon}_k)\right] \\
      =& -2\mathbb{E} \left[ \bm{H}^{\top}_k(\bm{x}_k)\bm{R}_k^{-1} (\bm{y}_k - \bm{h}_k(\bm {x}_k)) \right]  \\
        \approx& -\frac{2}{S}\sum_{s = 1} ^{S} \bm{H}^{\top}_k(\bm{g}(\bm{\epsilon}_k^s))\mathbb{E}[\bm{R}_k^{-1}](\bm{y}_k - \bm{h}_k(\bm{\epsilon}_k^s)),
    \end{split}
\end{equation}
and 
\begin{equation}
\label{RP_non-conjugate_P}
\begin{split}
    \nabla_{\bm{P}_{k|k}}\mathcal{D}(\bm{\mu}_k^{\bm{x}}) = & \mathbb{E}\left[ \nabla_{\bm{g}(\bm{\epsilon}_k)}\varphi(\bm{x}_k) \nabla_{\bm{P}_{k|k}}\bm{g}(\bm{\epsilon}_k)  \right]\\
    \approx & -\frac{1}{S}\sum_{s = 1} ^{S} \bm{H}^{\top}_k(\bm{g}(\bm{\epsilon}_k^s))\mathbb{E}[\bm{R}_k^{-1}] \\
    &\times(\bm{y}_k - \bm{h}_k(\bm{g}(\bm{\epsilon}_k^s)))
    (\bm{\epsilon}_k^{s})^{\top} \bm{L}_k^{-1}. 
\end{split}
\end{equation}

Based on the stochastic mirror descent method,  the natural parameters of $\bm{x}_k$ can be updated iteratively as
\begin{equation}
\label{total_natural_parameter_x}
\begin{split}
     (-0.5\bm{P}_{k|k}^{-1})^{(r+1)} = &  (-0.5\bm{P}_{k|k}^{-1})^{(r)} + \beta_k^{(r)} \nabla_{\bm{\mu}_k^{\bm{x}}{(2)}} \mathcal{\bar B}(\bm{\mu}_k^{\bm{x}}),
     \\
      (\bm{P}_{k|k}^{-1} \hat{\bm{x}}_{k|k})^{(r+1)} = &  (\bm{P}_{k|k}^{-1} \hat{\bm{x}}_{k|k})^{(r)} + \beta_k^{(r)} \nabla_{\bm{\mu}_k^{\bm{x}}(1)} \mathcal{\bar B}(\bm{\mu}_k^{\bm{x}}),
    \end{split}
\end{equation}
where $\beta_k^{(r)}$ is learning rate and $r$ is the iteration number.

Substituting the conjugate part in \eqref{conjugate_gradient} and the non-conjugate part in \eqref{RP_non-conjugate_x}-\eqref{RP_non-conjugate_P} into \eqref{total_natural_parameter_x}, yields
\begin{equation}
\label{update_invP}
    \begin{split}
        (\bm{P}_{k|k}^{-1})^{(r+1)} =& (1 - \beta_k^{(r)}) (\bm{P}_{k|k}^{-1})^{(r)} \\
        &+  \beta_k^{(r)} \left(\mathbb{E}[\bm{P}_{k|k-1}^{-1}] + \nabla_{\bm{P}_{k|k}} \mathcal{D}(\bm{\mu}_k^{\bm{x}}) \right ),
    \end{split}
\end{equation}
\begin{equation}
\label{update_x}
    \begin{split}
         \hat{\bm{x}}_{k|k}^{(r+1)} =& \hat{\bm{x}}_{k|k}^{(r)} + \beta_k^{(r)}  \bm{P}_{k|k} ^{(r+1)}  
        \mathbb{E}[\bm{P}_{k|k-1}^{-1}](\hat{\bm{x}}_{k|k-1} - \hat{\bm{x}}_{k|k}^{(r)}) \\
        &-0.5\beta_k^{(r)}  \bm{P}_{k|k} ^{(r+1)} \nabla_{\hat{\bm{x}}_{k|k}} \mathcal{D}(\bm{\mu}_k^{\bm{x}}).
    \end{split}
\end{equation}

An issue with the above updates \eqref{update_invP}-\eqref{update_x} is that the constraint is not taken into account. For multivariate Gaussian distribution, the covariance matrix $\bm{P}_{k|k}$ should be real-value and positive-definite. However, the learning rate $\beta_k^{(r)}$ in \eqref{update_invP} does not ensure that $\bm{P}_{k|k}$ are always positive. Setting the learning rate $\beta_k^{(r)}$ too high can cause the CVIAKF algorithm to diverge, and too low makes it slow to converge.

To address this issue, one can use the line-search approach \cite{lin2020handling}. The constraint can be satisfied by adding a compensation term into Eq.\eqref{update_invP}, that is:
   \begin{equation}
   \begin{split}
   \label{positive_definite}
   (\bm{P}_{k|k}^{-1})^{(r+1)} =& \frac{(\beta_k^{(r)})^2}{2} \hat{\bm{\mathrm{G}}}_k \bm{P}^{(r)}_{k|k} \hat{\bm{\mathrm{G}}}_k + (1 - \beta_k^{(r)}) (\bm{P}_{k|k}^{-1})^{(r)} \\
        &+  \beta_k^{(r)} \left(\mathbb{E}[\bm{P}_{k|k-1}^{-1}] + \nabla_{\bm{P}_{k|k}} \mathcal{D}(\bm{\mu}_k^{\bm{x}}) \right )
    \end{split}
   \end{equation}
   with 
   \begin{equation}
       \hat{\bm{\mathrm{G}}}_k = (\bm{P}_{k|k}^{-1})^{(r)} - \left(\mathbb{E}[\bm{P}_{k|k-1}^{-1}] +\nabla_{\bm{P}_{k|k}} \mathcal{D}(\bm{\mu}_k^{\bm{x}}) \right ).
   \end{equation}
The update \eqref{positive_definite} can substantiate that $\bm{P}_{k|k}^{-1}$ is always positive-definite~\cite{lin2020handling}.

\emph{2. Derivations of $\bm{\lambda}_k^{\bm{P}}$}: the update of natural parameter $\bm{\lambda}_k^{\bm{P}}$ is same as \eqref{eq::update hyper-parameter of P} since the measurement function $\bm{h}_k(\bm{x}_k)$ has no relationship with \eqref{B_P}.

\emph{3. Derivations of $\bm{\lambda}_k^{\bm{R}}$}: the update of natural parameter $\bm{\lambda}_k^{\bm{R}}$ remains to be \eqref{eq::update hyper-parameter of R}, whereas the term $\bm{A}_k$ is modified as
\begin{equation}
\begin{split}
    \bm{A}_k = \,&\mathbb{E}[(\bm{y}_k-\bm{h}_k(\bm{x}_k))(\bm{y}_k-\bm{h}_k(\bm{x}_k))^{\top}]
    \\
    \approx \,& \frac{1}{S}\sum_{s=1}^{S}[(\bm{y}_k-\bm{h}_k(\hat{\bm{x}}_{k|k} + \bm{L}_k \bm{\epsilon}_k^s))(\cdot)^{\top}].
    \end{split}
\end{equation}
The proof of \textbf{Theorem 2} has been completed. $\hfill \Large\square$

\bibliographystyle{IEEEtran} %the bibliographystyle
\bibliography{IEEEabrv,Reference} %the references
\end{document}